\documentclass[prx,twocolumn,aps,a4paper,superscriptaddress]{revtex4-1}
\usepackage[utf8]{inputenc} 
\usepackage{graphicx}

\usepackage{graphicx} 

\usepackage{color}
\usepackage{subfigure}
\usepackage{braket}
\usepackage{amsmath}
\usepackage{amssymb}
\usepackage{verbatim} 

\def\ii{{\rm i}}  \def\ee{{\rm e}}  
\def\be{\begin{equation}}
\def\ee{\end{equation}}
\def\bea{\begin{eqnarray}}
\def\eea{\end{eqnarray}}

\begin{document}

\title{Quantum dynamics of propagating photons with strong interactions: a generalized input-output formalism}
\author{Tommaso Caneva}
\affiliation{ICFO-Institut de Ciencies Fotoniques, Mediterranean Technology Park, 08860 Castelldefels (Barcelona), Spain}
\author{Marco T. Manzoni}
\affiliation{ICFO-Institut de Ciencies Fotoniques, Mediterranean Technology Park, 08860 Castelldefels (Barcelona), Spain}
\author{Tao Shi}
\affiliation{Max-Planck-Institut f\"ur Quantenoptik, Hans Kopfermann-strasse 1, 85748 Garching, Germany}
\author{James S. Douglas}
\affiliation{ICFO-Institut de Ciencies Fotoniques, Mediterranean Technology Park, 08860 Castelldefels (Barcelona), Spain}
\author{J. Ignacio Cirac}
\affiliation{Max-Planck-Institut f\"ur Quantenoptik, Hans Kopfermann-strasse 1, 85748 Garching, Germany}
\author{Darrick E. Chang}
\affiliation{ICFO-Institut de Ciencies Fotoniques, Mediterranean Technology Park, 08860 Castelldefels (Barcelona), Spain}

\begin{abstract}
There has been rapid development of systems that yield strong interactions between freely propagating photons in one dimension via controlled coupling to quantum emitters. This raises interesting possibilities such as quantum information processing with photons or quantum many-body states of light, but treating such systems generally remains a difficult task theoretically. Here, we describe a novel technique in which the dynamics and correlations of a few photons can be exactly calculated, based upon knowledge of the initial photonic state and the solution of the reduced effective dynamics of the quantum emitters alone. We show that this generalized ``input-output'' formalism allows for a straightforward numerical implementation regardless of system details, such as emitter positions, external driving, and level structure. As a specific example, we apply our technique to show how atomic systems with infinite-range interactions and under conditions of electromagnetically induced transparency enable the selective transmission of correlated multi-photon states.
\end{abstract}
\maketitle

\section{Introduction}
Systems in which individual photons can interact strongly with each other constitute an exciting frontier for the fields of quantum and nonlinear optics \cite{NLO14}. Such systems enable the generation and manipulation of non-classical light, which are crucial ingredients for quantum information processing and quantum networks \cite{K08}. At the many-body level, it has been predicted that these systems can produce phenomena such as quantum phase transitions of light \cite{HAR06,GRE06} or photon crystallization \cite{CRY08,OTT13}. Early examples of such systems where strong interactions between photons could be observed consisted of individual atoms coupled to single modes of high-finesse optical cavities, within the context of cavity quantum electrodynamics~(QED) \cite{TUR95,BIR05}. More recently, a number of systems have emerged that produce strong optical nonlinearities between freely propagating photons, including cold atomic gases coupled to guided modes of tapered fibers \cite{VET10,GOB12}, photonic crystal fibers \cite{BAJ09} or waveguides \cite{GOB14}, cold Rydberg gases in free space \cite{rydberg_experiments}, and superconducting qubits coupled to microwave waveguides \cite{HOI13,LOO13}.

\begin{figure}
\includegraphics[width=85mm,angle=0,clip]{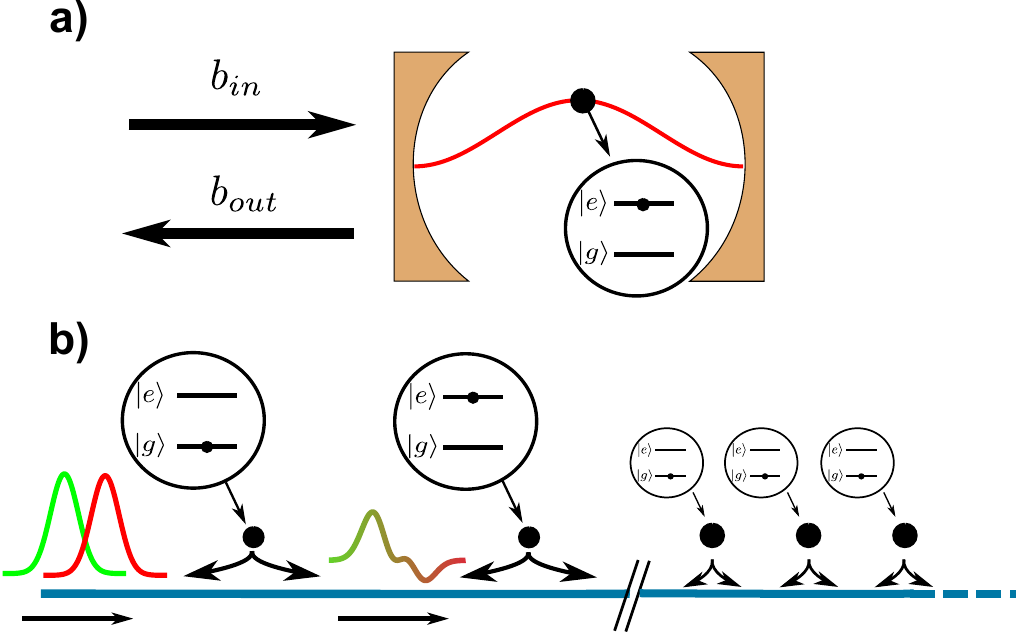}
\caption{{\bf Cavity QED vs. many-atom waveguide} (a) Schematic representation of a cavity QED system. The input-output formalism enables the outgoing field to be calculated based on knowledge of the input field and the internal dynamics of the cavity system. (b) System composed of many atoms coupled to a common waveguide. The nonlinearity of an atom enables the generation of a continuum of new frequencies upon scattering of an incoming state. This property, combined with multiple scattering from other atoms, appears to make this system more complicated than the cavity QED case.}
\label{fig_1}
\end{figure}

The paradigm of cavity QED admits exact analytical or numerical solutions at the few-photon~(and/or few-atom) level through the elegant ``input-output'' formalism \cite{GAD95}, as illustrated conceptually in Fig.~\ref{fig_1}(a). In this formalism, all of the properties of the field exiting the system (the output) can be determined based upon knowledge of the input field and the dynamics of the atom-cavity system alone. For a few excitations, the latter can be solved due to the small Hilbert space associated with a small number of excitations of a discrete cavity mode. In contrast, despite rapid development on the experimental front, theoretical techniques to treat the dynamics of freely propagating photons interacting with spatially distributed emitters are generally lacking. At first glance, the challenge compared to the cavity case arises from the fact that a two-level system is a nonlinear frequency mixer, which is capable of generating a continuum of new frequencies from an initial pulse, as schematically depicted in Fig.~\ref{fig_1}(b). A priori, keeping track of this continuum as it propagates and re-scatters from other emitters appears to be a difficult task. An exception is the weak excitation limit, in which atoms can be treated as linear scatterers and the powerful transfer matrix method of linear optics can be employed~\cite{DEC12,DEU95}.

The full quantum case has been solved exactly in a limited number of situations in which nonlinear systems are coupled to 1D waveguides, such as up to three photons scattering on a single atom~\cite{SHI09,FAN10}, one and two photons scattering on a cavity QED system \cite{SHI11}, and many photons scattering on many atoms coupled to a chiral (monodirectional) waveguide \cite{PLE12}. The formalism employed in Ref.~\cite{FAN10} is particularly elegant, because it establishes an input-output relation to determine the nonlinear scattering from a two-level atom. Here, we show that this technique can be efficiently generalized to many atoms, chiral or bi-directional waveguides, and arbitrary atomic configurations, providing a powerful tool to investigate nonlinear optical dynamics in all systems of interest.

This paper is organized in the following way: first, we present a generalized input-output formalism to treat few-photon propagation in waveguides coupled to many atoms. We show that the infinite degrees of freedom associated with the photonic modes can be effectively integrated out, yielding an open, interacting ``spin'' model that involves only the internal degrees of freedom of the atoms. This open system can be solved using a number of conventional, quantum optical techniques. Then, we show that the solution of the spin problem can be used to re-construct the optical fields. In particular, we provide a prescription to map spin correlations to S-matrix elements, which contain full information about the photon dynamics, and give explicit closed-form expressions for the one- and two-photon cases. Importantly, in analogy with the cavity QED case, our technique enables analytical solutions under some scenarios, but in general allows for simple numerical implementation under a wide variety of circumstances of interest, such as different level structures, external driving, atomic positions, atomic motion, etc. Finally, to illustrate the ease of usage, we apply our technique to the study of nonlinear field propagation through an optically dense ensemble of atoms with Rydberg-like interactions.

\section{Generalized input-output formalism}

In this section we consider a generic system composed of many atoms located at positions $z_i$ along a bidirectional waveguide. We assume that there is an optical transition between ground and excited-state levels $\ket{g}$ and $\ket{e}$ to which the waveguide couples, but otherwise we leave unspecified the atomic internal structure and the possible interactions between them (\textit{e.g.}, Rydberg interactions), as such terms do not affect the derivation presented here. The bare Hamiltonian of the system is composed of a term describing the energy levels of the atoms $H_\text{at}$, and a waveguide part $H_\text{ph} = \sum_{v = \pm}\int dk \,\omega_k b^\dagger_{v,k}b_{v,k}$, where $k$ is the wavevector and $v=\pm$ is an index for the direction of propagation, with the plus (minus) denoting propagation towards the right (left) direction. We assume that within the bandwidth of modes to which the atoms significantly couple, the dispersion relation for the guided modes can be linearized as $\omega_k=c|k|$. The interaction between atoms and photons is given by
\begin{equation}
H_\text{int} = g\sum_{v = \pm}\sum_{i=1}^N \int dk\,\big(b_{v,k}\sigma_{eg}^i\,e^{\ii vkz_i} + \text{h.c.}\big),
\end{equation}
which describes the process where excited atoms can emit photons into the waveguide, or ground-state atoms can become excited by absorbing a photon. The coupling amplitude $g$ is assumed to be identical for all atoms, while the coupling phase depends on the atomic position ($e^{\ii vkz_i}$). Here, we will explicitly treat the more complicated bidirectional case, although all of the results readily generalize to the case of a single direction of propagation.

Our immediate goal is to generalize the well-known input-output formalism of cavity QED~\cite{GAD95} and the more recent formalism for a single atom coupled to a waveguide \cite{FAN10}, to the present situation of many spatially distributed atoms coupled to a common waveguide. In short, we will eliminate the photonic degrees of freedom by formal integration, which reveals that the output field exiting the collection of atoms is completely describable in terms of the input field and atomic properties alone. This formal integration also provides a set of generalized Heisenberg-Langevin equations that governs the atomic evolution.

The Heisenberg equations of motion for $\sigma_{ge}^i$ and $b_{v,k}$ can be readily obtained by calculating the commutators with $H$. To simplify the presentation of the resulting equations, we replace the spin operators $\sigma_{ge}^i$ with the bosonic annihilation operator $a_i$. The two-level nature of the atomic transition can be retained by introducing an interaction energy $U_0$ for multiple excitations, through the term $(U_0/2)\sum_i a^\dagger_ia_i(a^\dagger_ia_i-1)$ in $H_\text{at}$, and by taking the limit $U_0 \rightarrow \infty$ at the end of calculations for observable quantities.

The Heisenberg equations for $b_{v,k}$ can be formally integrated and Fourier transformed, to obtain the real-space wave equation
\begin{eqnarray}
b_{v}(z,t)&=&b_{v,\text{in}}(t-vz/c)\nonumber\\
&-&\frac{\ii \sqrt{2\pi}g}{c}\,\sum_{j=1}^N\,\theta\big(v(z-z_j)\big)a_j\bigg(t-{\frac{v(z-z_j)}{c}}\bigg).
\label{eq:b}
\end{eqnarray}
Here $b_{v,\text{in}}$ is the homogeneous solution, physically corresponding to the freely propagating field in the waveguide, while the second term on the right consists of the part of the field emitted by the atoms. Inserting Eq.~\eqref{eq:b} into the equation for $a_i$, we obtain
\begin{eqnarray}
\label{eq:a_2}
\dot{a}_i&=&\ii[H_\text{at}, a_i] - \ii\sqrt{2\pi}g\sum_{v = \pm}\,b_{v,\text{in}}(t-vz_i/c)\nonumber\\
&-&\frac{2\pi g^2}{c}\sum_{j=1}^N\,a_j(t-|z_i-z_j|/c)).
\end{eqnarray}

In realistic systems, time retardation can be neglected, resulting in the Markov approximation $a_j(t-|z_i-z_j|/c) \approx a_j(t)e^{\ii \omega_{\text{in}}|z_i-z_j|/c}$. Here, $\omega_{\text{in}}$ is a central frequency around which the atomic dynamics is centered (typically the atomic resonance frequency). This approximation is valid when the difference in free-space propagation phases $ \Delta\omega L/c \ll 1$ is small across the characteristic system size $L$ and over the bandwidth of photons $\Delta\omega$ involved in the dynamics. As a simple example, the characteristic bandwidth of an atomic system is given by its spontaneous emission rate, corresponding to a few MHz, which results in a significant free-space phase difference only over lengths $L \gtrsim 1$ m much longer than realistic atomic ensembles.

We have thus obtained the generalized Heisenberg-Langevin equation
\begin{eqnarray}
\label{eq:many_1}
\dot{a}_i&=&\ii[H_\text{at}, a_i]  - \ii\sqrt{c\,\Gamma_{1D}/2}\sum_{v = \pm}\,b_{v,\text{in}}(t-vz_i/c)\nonumber\\
&-&\frac{\Gamma_{1D}}{2}\sum_{j=1}^N\,a_j\,e^{\ii \omega_\text{in}|z_i-z_j|/c},
\end{eqnarray}
where we have identified $\Gamma_{1D} = 4\pi g^2/c$ as the single-atom spontaneous emission rate  into the waveguide modes. If we keep separated the terms proportional to $a_j$ coming from the right and left-going photonic fields, we can find easily that the Lindblad jump operators corresponding to the decay of the atoms into the waveguide are $O_\pm = \sqrt{\Gamma_{1D}/4}\sum_j\,a_j e^{\mp\ii \omega_\text{in}z_j/c}$, in terms of which we can write the master equation for the atomic density matrix $\dot{\rho} = \mathcal{L}[\rho] \equiv -\ii[H_\text{at}, \rho] + \sum_{v=\pm}\,2O_v\rho O^\dagger_v - O^\dagger_vO_v\rho - \rho O^\dagger_v O_v$.
We also see that we can derive Eq.~\eqref{eq:many_1} from a non-Hermitian effective Hamiltonian
\begin{equation}
H_\text{eff} = H_{\text{at}} - \frac{\ii\Gamma_{1D}}{2}\sum_{ij}\,a_i^\dagger a_j\,e^{\ii \omega_\text{in}|z_i-z_j|/c},
\label{eff_H:eq}
\end{equation}
which can be used for a quantum jump description of the atomic dynamics. The resulting infinite-range interaction between a pair of atoms $i$,$j$ intuitively results from the propagation of a mediating photon between that pair, with a phase factor proportional to the separation distance.

Within the same approximations employed above to derive the Heisenberg-Langevin equations we can obtain a generalized input-output relation of the form $b_{v,\text{out}}(z,t) = b_{v,\text{in}}(t-vz/c) - \ii\sqrt{\Gamma_{1D}/(2c)}\sum_{i=1}^N\,a_i(t)\,e^{\ii v\omega_\text{in}(z-z_i)/c}$, where the output field is defined for $z > z_{R} \equiv \max[z_i]$ ($z < z_L \equiv \min[z_i]$) for right(left)-going fields. However, since the right-going output field propagates freely after $z_R$, it is convenient to simply define $b_{+,\text{out}}(t) = b_{+,\text{out}}(z_R+\epsilon,t)$ as the field immediately past the right-most atom (where $\epsilon$ is an infinitesimal positive number), and similarly for the left-going output. The derived relation shows that the out-going field properties are obtainable from those of the atoms alone.

The emergence of infinite-range interactions between emitters mediated by guided photons, and input-output relationships between these emitters and the outgoing field, have been discussed before in a number of contexts \cite{DEC12,FAN10,LUM13}, but the idea that such concepts could be used to study quantum interactions of photons in extended systems has not been fully appreciated. In the remaining sections, we will demonstrate the effectiveness of this approach to quantum nonlinear optics. In particular, the infinite-dimensional continuum of the photons is effectively reduced to a Hilbert space of dimension $\text{dim}[\mathcal{H}] = \sum^n_{i=0}\binom {N} {i}$ where $n$ is the maximum number of atomic excitations (for $n = N$ we have $\text{dim}[\mathcal{H}] = 2^N$). The atomic dynamics, on the other hand, having been reduced to standard Heisenberg-Langevin equations, quantum jump, or master equations, are solvable by conventional prescriptions~\cite{meystre}. It is also possible to derive generalized master equations describing the atomic dynamics in response to arbitrary (\textit{e.g.}, non-classical) incident states of light, as detailed in Appendix C.

\section{Relation to S-matrix elements}

The S-matrix characterizes how an incoming state of monochromatic photons evolves via interaction with atoms into a superposition of outgoing monochromatic photons. Because monochromatic photons form a complete basis, the S-matrix thus contains all information about photon dynamics. Here, we will show that S-matrix elements can be calculated from the dynamics of the atoms evolving under the effective spin model. 

Formally, the S-matrix for the interaction of an $n$-photon state with an arbitrary system is defined as
\begin{eqnarray}
\label{eq:s_1}
S^{(n)}_{\mathbf{p};\mathbf{k}}&=&\bra{\mathbf{p}} S\ket{\mathbf{k}}\nonumber \\
&=&\bra{0}b_\text{out}(p_1)..b_\text{out}(p_n)b^\dagger_\text{in}(k_1)..b^\dagger_\text{in}(k_n)\ket{0}\nonumber \\
&=&\mathcal{FT}^{(2n)}\bra{0}b_\text{out}(t_1)..b_\text{out}(t_n)b^\dagger_\text{in}(t_1')..b^\dagger_\text{in}(t_n')\ket{0},
\end{eqnarray}
where the input and output creation operators respectively create freely propagating incoming and outgoing photonic states. The vectors $\mathbf{p}$ and $\mathbf{k}$ denote the outgoing and incoming frequencies of the $n$ photons. The input and output operators can be any combination of + and - propagation directions (we have omitted this index here for simplicity). In the last line we have used a global Fourier transformation $\mathcal{FT}^{(2n)} = (2\pi)^{-n}\int\prod_{i=1}^n\,dt_idt_i'\,e^{\ii c(t_ip_i-t_i'k_i)}$, to express the S-matrix in time.

The S-matrix has been calculated exactly before in a limited number of situations \cite{SHI09,FAN10,SHI11,PLE12}. Here, we provide a general prescription to numerically obtain the S-matrix starting from the input-output formalism. First, it should be noted that the operators $b_\text{in}$, $b_\text{out}$ correspond to the input and output operators defined in the previous section \cite{FAN10}. On the other hand, the input-output relation enables the correlator of Eq. \eqref{eq:s_1} to be written purely in terms of atomic operators. For notational simplicity, we give a derivation for a single spin and a monodirectional waveguide, but its generalization to the bidirectional waveguide and many atoms is straightforward. For our purpose it is enough to have an input-output relation of the form
\begin{equation}
\label{eq:bound_1}
b_{\text{out}} = b_{\text{in}} - \ii\sqrt{\gamma}a,
\end{equation}
where $a$ is in our case the spin operator $\sigma_{ge}$.

We summarize the main idea of the derivation here, while the details can be found in Appendix A.1. We begin by noting that Eq.~(\ref{eq:bound_1}) enables one to replace output operators by a combination of system and input operators, or input operators by system and output operators. Selectively using these substitutions, one can exploit favorable properties of either the input or output field, in order to gradually time order all of the system operators~(where operators at later times appear to the left of those at earlier times), while removing input and output operators from the correlation. The favorable properties that can be used are that 1) system and input operators commute, $[a(t), b_\text{in}(t')] = 0$, when $t'>t$, and likewise  $[a(t), b_\text{out}(t')] = 0$ for $t' < t$, 2) input annihilation operators at different times commute amongst themselves, as do output annihilation operators, and 3) the correlator can be reduced in size using $[b_\text{in}(t),b^{\dagger}_\text{in}(t')]=\delta(t-t')$, or made to vanish using $b_\text{in}(t)\ket{0}=0$ or $\bra{0}b^{\dagger}_\text{out}(t)=0$. Through this procedure, the S-matrix can be expressed as a Fourier transform of a sum of terms involving only time-ordered atomic operators~(indicated by the operator $T$). These functions generally have the form
\begin{equation}
\label{eq:s_7}
\bra{0}T\big[a(t_1)..a(t_m)a^\dagger(t_1')..a^\dagger(t_m')\big]\ket{0},
\end{equation}
with $m \leq n$, multiplied by $n-m$ delta functions in time.

Moreover, using the general expression for the Heisenberg-Langevin equation and the quantum regression theorem, it can be proven that when external fields driving the system do not generate waveguide photons, the correlation function of Eq. \eqref{eq:s_7} can be evaluated by evolving $a(t)$ as $e^{\ii H_\text{eff}t}\,a\,e^{-\ii H_\text{eff}t}$ (see Appendix A.2 for a formal derivation). Here, $H_\text{eff}$ is the effective Hamiltonian from Eq. \eqref{eff_H:eq} that contains only the spin operators. Although such a form for $a(t)$ is not true in general due to quantum noise, these noise terms have no influence on the correlation. A similar procedure as above enables one to express other important observables of the field, such as the second-order correlation function $g^{(2)}(t)$, in terms of correlation functions involving atoms alone.

While the discussion has thus far been completely general, the case of S-matrix elements involving only one or two photons can be formally reduced to particularly simple expressions. For example, in Appendix B, we show that the transmission coefficient $T_k$ for the many-atom, bi-directional waveguide case is related to the S-matrix by $S^{(1)}_{p+;k+}\equiv \bra{0}b_{+,\text{out}}(p)b^{\dagger}_{+,\text{in}}(k)\ket{0}\equiv T_k\delta_{pk}$. Furthermore, it can be expressed in terms of a known $\sim N\times N$ matrix corresponding to the single-excitation Green's function $G_0$~(whose form varies depending on the system details),
\begin{equation}
T_k=1-\frac{\ii\Gamma_{1D}}{2}\sum_{ij}\left[G_0(k)\right]_{ij}e^{-\ii k_\text{in}(z_i-z_j)}.
\end{equation}
Similarly, the two-photon S-matrix in transmission is generally given by
\begin{multline}
S^{(2)}_{p_1+,p_2+;k_1+,k_2+}=T_{k_1}T_{k_2}\delta_{p_1 k_1}\delta_{p_2 k_2} + \\ - \ii\frac{\Gamma^2_{1D}}{8\pi}\delta_{p_1+p_2,k_1+k_2}\sum_{iji'j'}W_{ij}T_{ij;i'j'}W_{i'j'}+(p_1\leftrightarrow p_2),\label{eq:S2}
\end{multline}
where the first and second terms on the right describe the linear and nonlinear contributions, respectively. The latter term can be expressed in terms of matrices $W$ related to the single-excitation Green's function, and a known $\sim N^2 \times N^2$ matrix $T$ characterizing atomic nonlinearities and interactions.

\section{Electromagnetic Induced Transparency}
\begin{figure}[t]
	\setlength{\unitlength}{1cm}
	\begin{picture}(8,4.5)
	\put(-0.3,0){\raisebox{4ex}{
	\includegraphics[width=25mm,angle=0,clip]{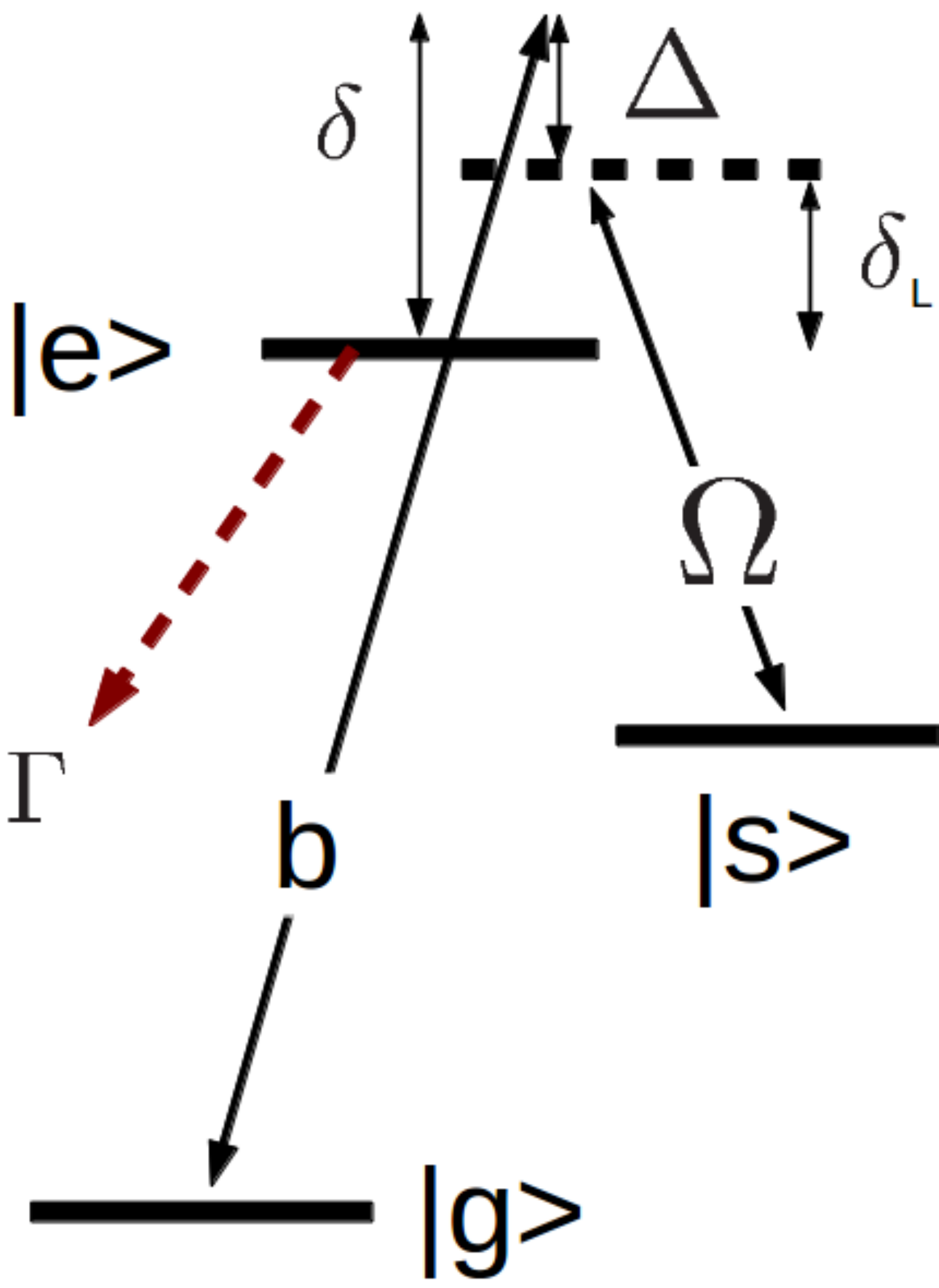}}}
	\put(0,4){(a)}
	\put(3,0) {\includegraphics[width=53mm,angle=0,clip]{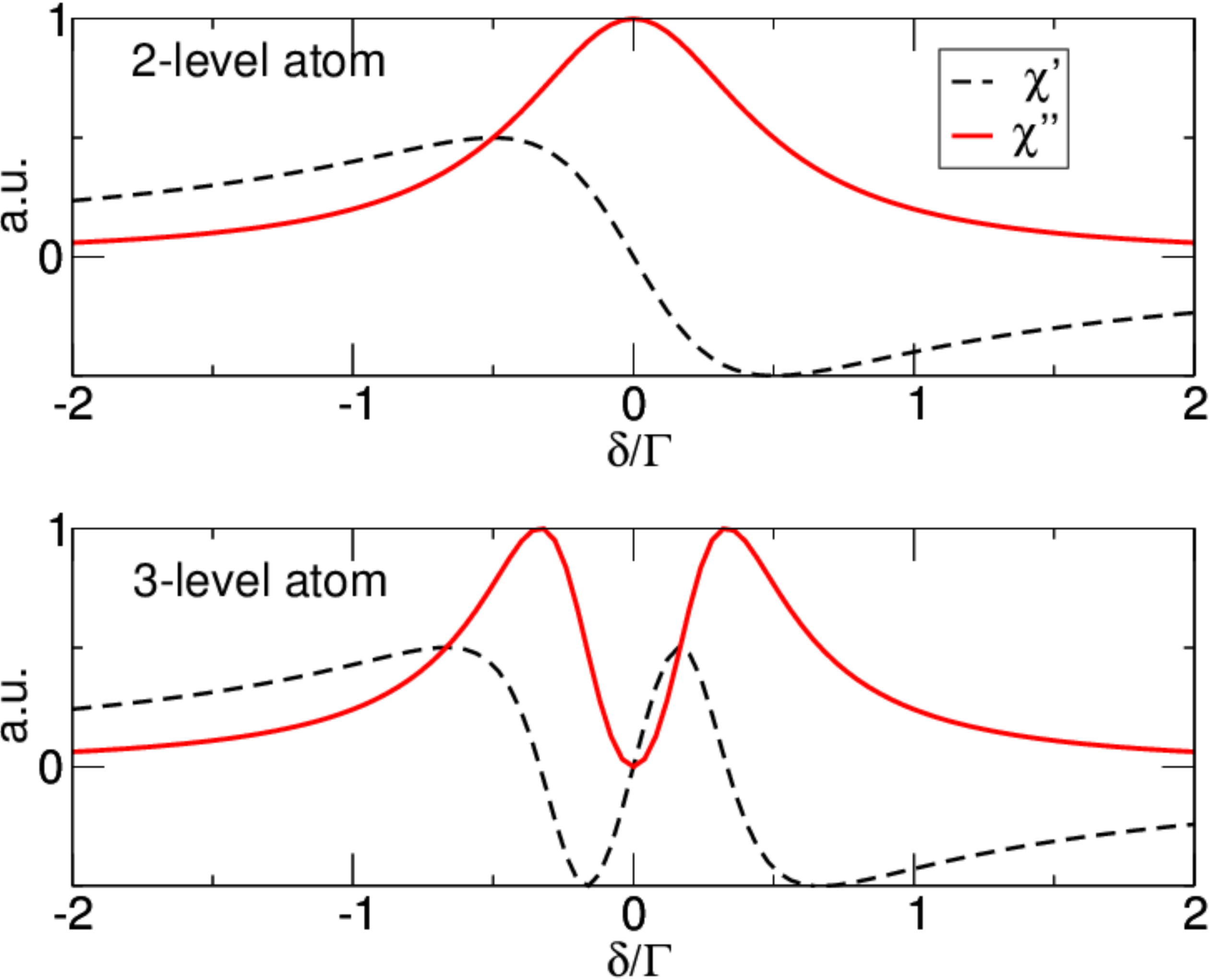}}
	\put(2.6,4){(b)}
	\end{picture}
\caption{(a) EIT level scheme. The atomic ground~($\ket{g}$) and excited states~($\ket{e}$) interact with the quantum propagating field~$b$ of the waveguide. An additional classical field with Rabi frequency $\Omega$ couples state $\ket{e}$ to a metastable state $\ket{s}$. The total single-atom linewidth of the excited state is given by $\Gamma$. (b) The real ($\chi '$) and imaginary
($\chi ''$) parts of the linear susceptibility for a two-level atom (upper panel) and three-level atom
(lower panel), as a function of the dimensionless detuning $\delta/\Gamma$ of the field $b$ from the resonance frequency of the $\ket{g}$-$\ket{e}$ transition. For the three-level atom, the parameters used are $\delta_L=0$ and $\Omega/\Gamma=1/3$.}
\label{eit_level:fig}
\end{figure}
\begin{figure}[t]
	\setlength{\unitlength}{1cm}
	\begin{picture}(8,4.5)
	\put(0.4,0){\includegraphics[width=65mm,angle=0,clip]{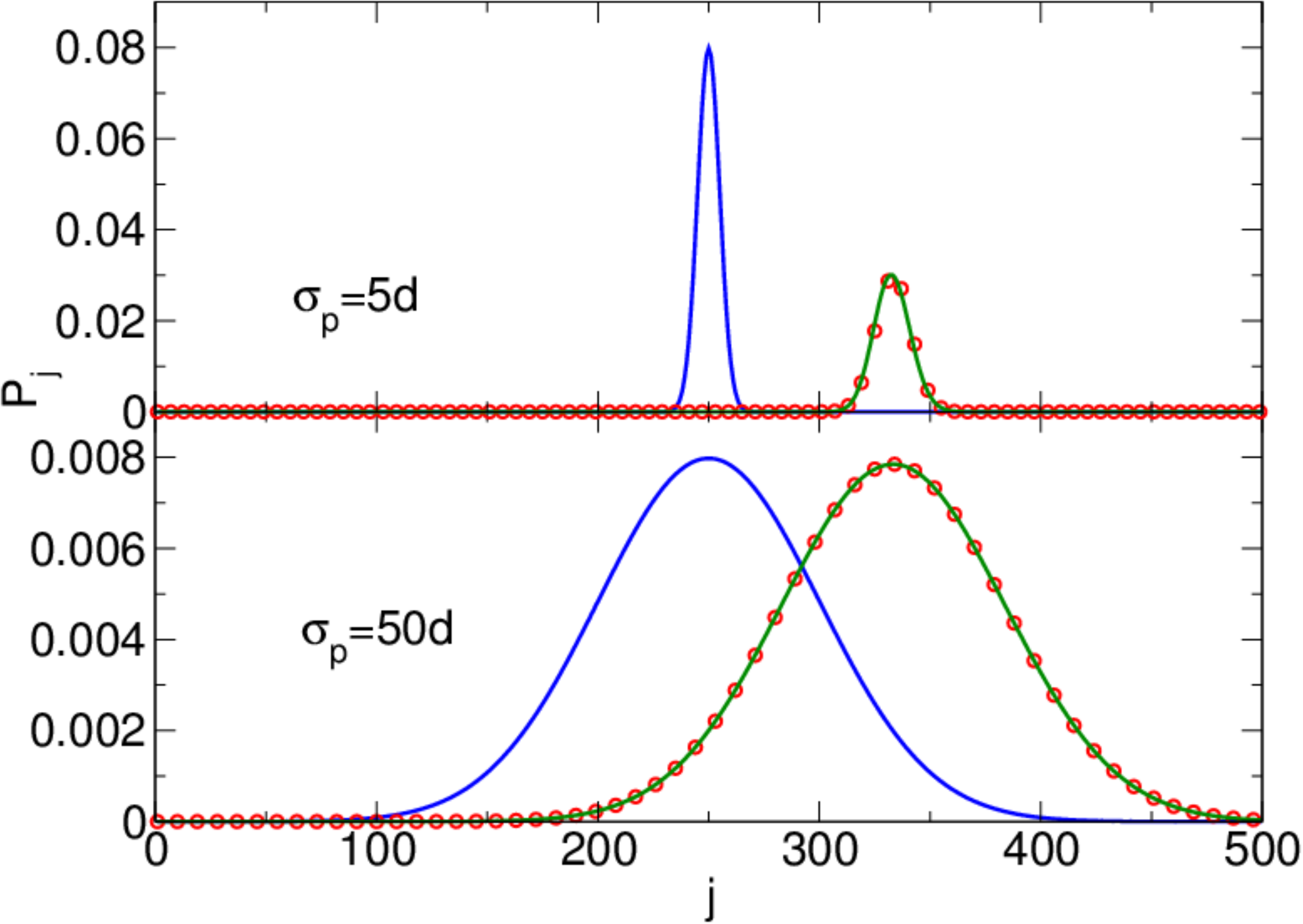}}
	\put(1.5,4){(a)}
	\put(1.5,2){(b)}
        \end{picture}
\caption{EIT: single polariton propagation for $\sigma _p<\sigma _{EIT}$ (a)
	and $\sigma _p >\sigma _{EIT}$ (b). Plotted is the population
	$P_j=\langle\sigma_{ss}^j\rangle$
	of atom $j$ in the state $|s\rangle$. The blue line corresponds to the initial state,
	the green line to the state evolved over a time $t_f$, and the red dots to the
	theoretically predicted evolution. 	
	Other parameters: $t_f v_g=Nd/6$, $\Gamma=5$, $\Gamma _{1D}=10$, $\Omega =1$, $N=500$ and
	$\sigma _{EIT}\sim 22d$, where $d$ is the lattice constant.}
\label{eitSinglePolariton:fig}
\end{figure}
In this section, we apply our formalism to a specific example involving three-level atoms under conditions of electromagnetically induced
transparency (EIT) and with Rydberg-like interactions between atoms ~\cite{rydberg_experiments}.
The linear susceptibility for a two-level atom with states $|g\rangle$ and $|e\rangle$,
in response to a weak probe field with detuning $\delta=\omega _p-\omega _{eg}$ from the atomic
resonance, is shown in Fig.~\ref{eit_level:fig}b. It can be seen that the response on resonance
is primarily absorptive, as characterized by the imaginary part of the susceptibility
($\chi''$, red curve). In contrast, the response can become primarily dispersive near resonance
if a third level $|s\rangle$ is added,
and if the transition $|e\rangle -|s\rangle$ is driven by a control
field~(characterized by Rabi frequency $\Omega$ and single photon detuning
$\delta _L =\omega _L-\omega _{es}$). Specifically, via interference between the probe and control
fields, the medium can become transparent to the probe field ($\chi ''=0$) when two-photon
resonance is achieved, $\delta-\delta_L=0$, realizing EIT~\cite{eit_papers}.
In this process, the incoming probe field strongly mixes with spin wave excitations $\sigma_{sg}$
to create ``dark-state polaritons."
The medium remains highly transparent within a characteristic bandwidth
$\Delta_{EIT}$ around the two-photon resonance, which reduces to
$\Delta _{EIT}\sim 2\Omega ^2/(\Gamma \sqrt{D}) $,
when $\delta_L=0$. Here we have introduced the total single-atom linewidth $\Gamma$ (see below) and the optical depth $D$, corresponding to the opacity of the medium in absence of EIT
and defined in terms of input and output intensity as
$I_{\rm out}(\Omega =0)/I_{\rm in}(\Omega =0)=\exp(-D)$. These polaritons propagate at a
strongly reduced group velocity $v_g \ll c$, as indicated by the steep slope of the
real part of the susceptibility $\chi'$ in Fig.~\ref{eit_level:fig}b, which is proportional to the control field
intensity~\cite{eit_papers}.

Taking $s_i$ to be the annihilation operator for the state
$|s_i\rangle$, EIT is described within our spin model by the effective spin Hamiltonian
\begin{eqnarray}
 H&=&-(\delta_L + \ii\frac{\Gamma'}{2})\sum _j a^\dagger _j a _j
-\Omega\sum _j (a^\dagger _js_j +s^\dagger _j a _j)\nonumber\\
&-& \ii\frac{\Gamma _{1D}}{2}\sum _{j,l}e^{\ii k_\text{in}|z_j-z_l|}a^\dagger _j a _l,
\label{eit_ham:eq}
\end{eqnarray}
where the first line represents the explicit form of $H_{at}$ in Eq.~(\ref{eff_H:eq})
for the EIT three-level atomic structure.
In addition to waveguide coupling, here we have added an independent atomic decay rate $\Gamma'$ into other channels (\textit{e.g.}, unguided modes), yielding a total single-atom linewidth of $\Gamma=\Gamma'+\Gamma_{1D}$.

Our theoretical model nearly ideally describes experiments in which atoms are coupled
to one-dimensional waveguides such as nanofibers~\cite{VET10} or photonic crystals~\cite{GOB12},
in which the interaction probability of a single photon and atom is
$\Gamma_{1D}/\Gamma\sim 0.1$ is significant and up to $N\sim 10^3$ atoms are trapped.
One consequence of the large interaction probability is that even a single atom can yield a significant reflectance for single photons~\cite{GOB12}. In free-space experiments,
the atom-photon interaction probability is much smaller, while a much larger number of
atoms is used to induce a significant optical response. While this large number cannot
be implemented numerically with our model, our system can nonetheless be used to reproduce
macroscopic observables, thus making our approach an excellent description for atom-light
interfaces in general.

Intuitively (and as can be rigorously shown, see below), provided that free-space
experiments only involve a single transverse mode of light, one expects that the system response
to light only depends on the interaction probability and number of atoms via their product.
This product in fact defines the optical depth $D=N(2\Gamma_{1D}/\Gamma)$~\cite{opticalDepth}.
As $D\lesssim 10^2$ in free-space experiments, such systems can be evaluated using our
spin model with several hundred atoms and suitably large $\Gamma_{1D}/\Gamma$.
The only remaining issue is the potentially large reflection generated by atoms in the
mathematical model, compared to free-space ensembles where reflection is negligible.
Reflection can be suppressed in the model by giving the atoms a lattice constant $d$
where $k_\text{in}d=(2m+1)\pi/2$, where $m$ is a non-negative integer, such that the reflection from atoms destructively
interferes~\cite{Chang_NJP11,perfectMirror}. All subsequent numerical calculations are performed under this condition.

The spin model on a lattice describing EIT, Eq.~(\ref{eit_ham:eq}), can be exactly solved
in the linear regime using the transfer matrix
formalism~\cite{Chang_NJP11}, which correctly
reproduces the free-space result and dependence on optical depth for
group velocity $v_g=2\Omega^2 n/\Gamma_{1D}$ and transparency window $\Delta_{EIT}$, where $n$ is the (linear) atomic density. The corresponding minimum spatial extent of a pulse that can propagate inside
the medium with high transparency is given by
$\sigma_{EIT}={v}_g/{\Delta}_{EIT}$.

A single photon propagating inside an ensemble of atoms under
EIT conditions is coherently mapped onto a single dark polariton,
corresponding to a delocalized spin wave populating
the single excitation subspace of the atomic ensemble. The polariton dynamics
can be therefore visualized directly by monitoring
the excitation probability $\langle \sigma _{ss}^j\rangle$ of the atoms in the ensemble.
In Fig.~\ref{eitSinglePolariton:fig}, we initialize a single polariton inside the medium
with an atomic wave function of the form
$|\psi\rangle=\sum_{j} f_j \sigma_{sg}^j |g\rangle^{\otimes N}$,
and we determine numerically the time evolution under $H$ in Eq.~(\ref{eit_ham:eq})
up to a final time $t_f$. Choosing an initially Gaussian spin wave, $f_j=\exp(\ii k_\text{in}dj)\exp(-(jd-\mu)^2/4\sigma _{p}^2)/(2\pi \sigma _p ^2)^{1/4}$,
with spatial extent
$\sigma _p$ (blue line), one sees that the wavepacket propagates a distance $v_g\cdot t_f$, and with
little loss provided that $\sigma _p>\sigma_{EIT}$. Numerics (green line) show perfect agreement with theoretical predictions
(red lines) obtained via the transfer matrix formalism~\cite{Chang_NJP11}.

\section{Infinite range interaction}

\begin{figure}[top]
	\setlength{\unitlength}{1cm}
	\begin{picture}(8,15)
	\put(0.5,10) {\includegraphics[width=60mm,angle=0,clip]{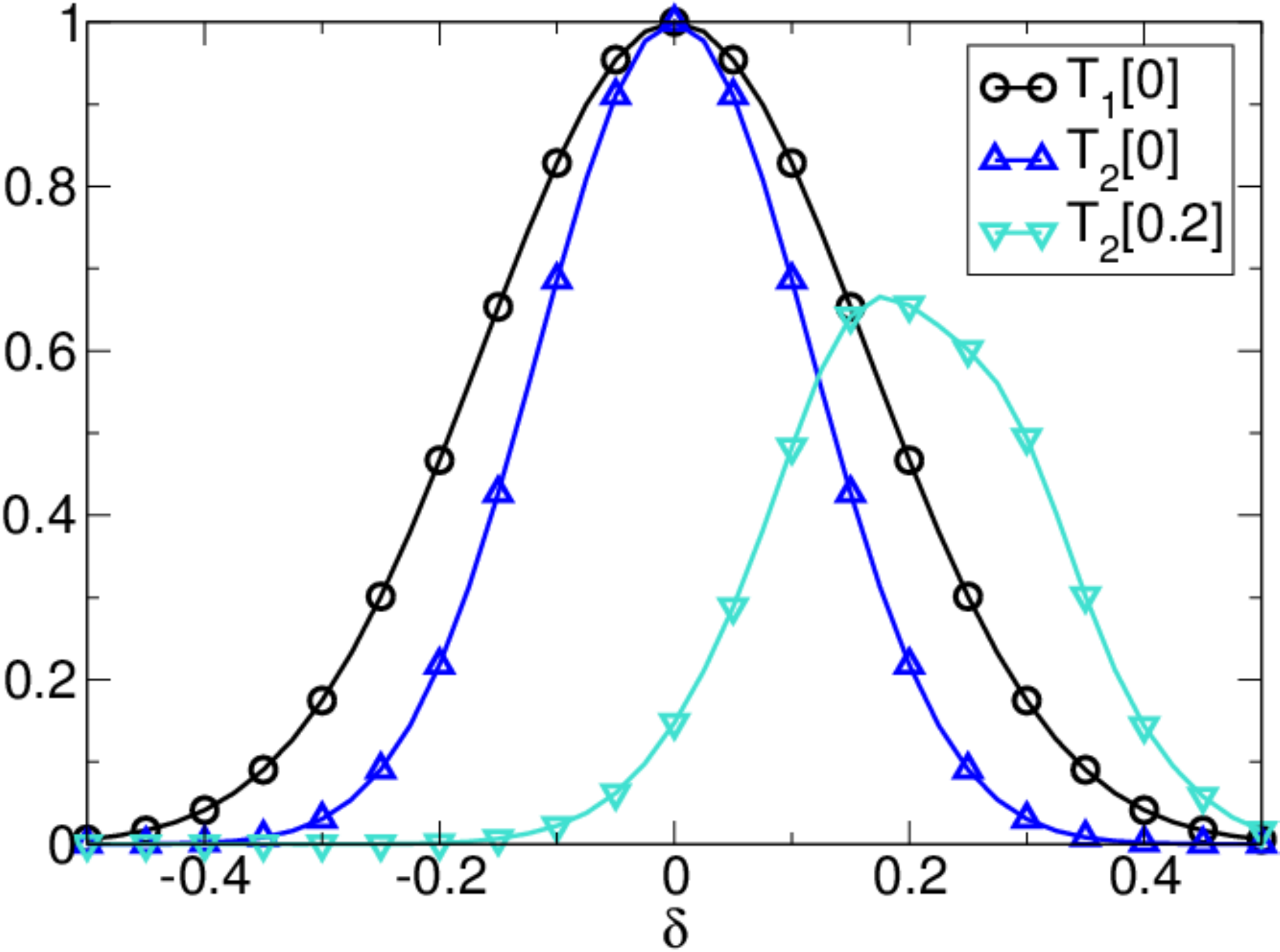}}
	\put(0,4.6) {\includegraphics[width=75mm,angle=0,clip]{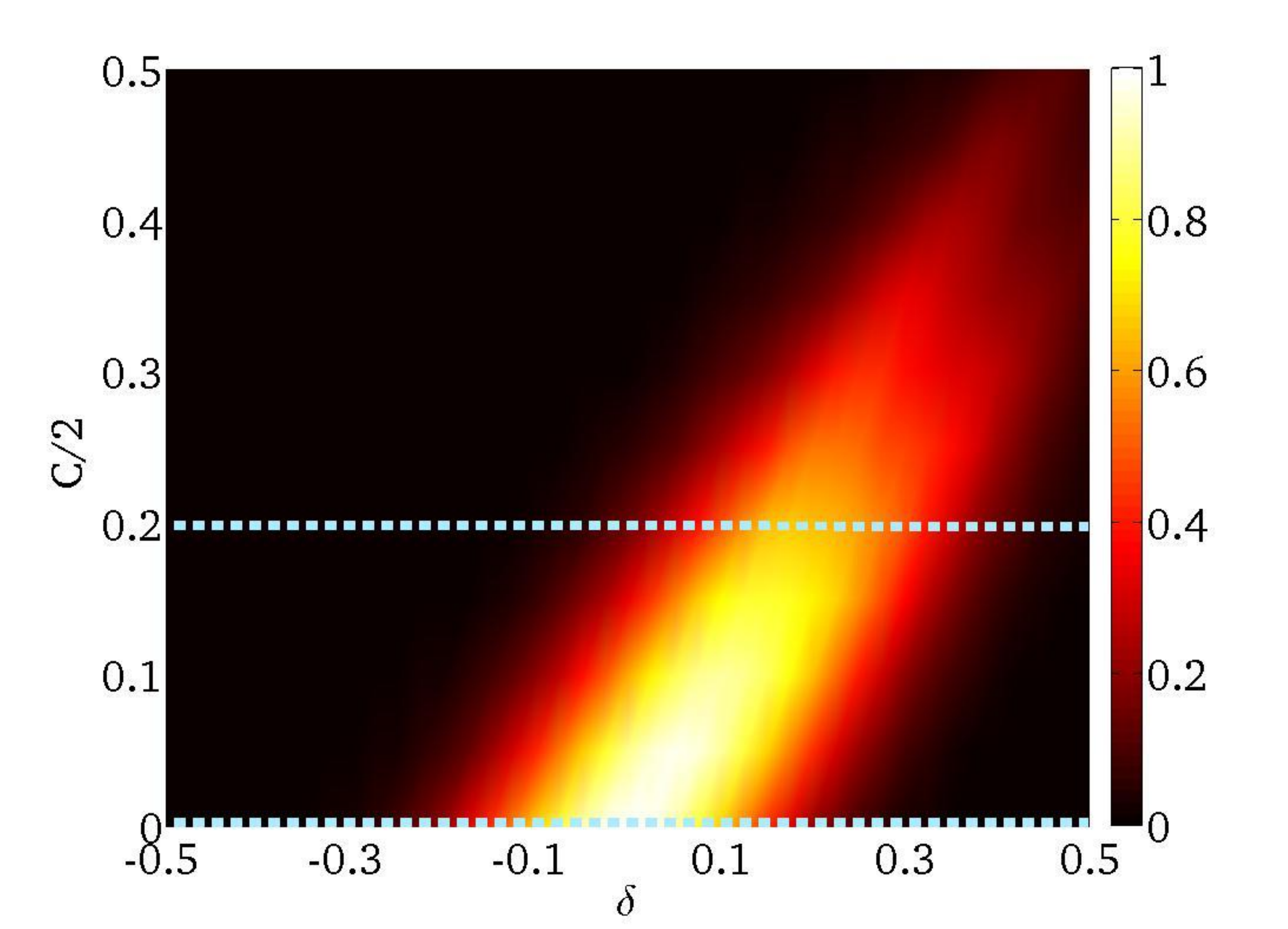}}
	\put(0.3,0) {\includegraphics[width=62mm,angle=0,clip]{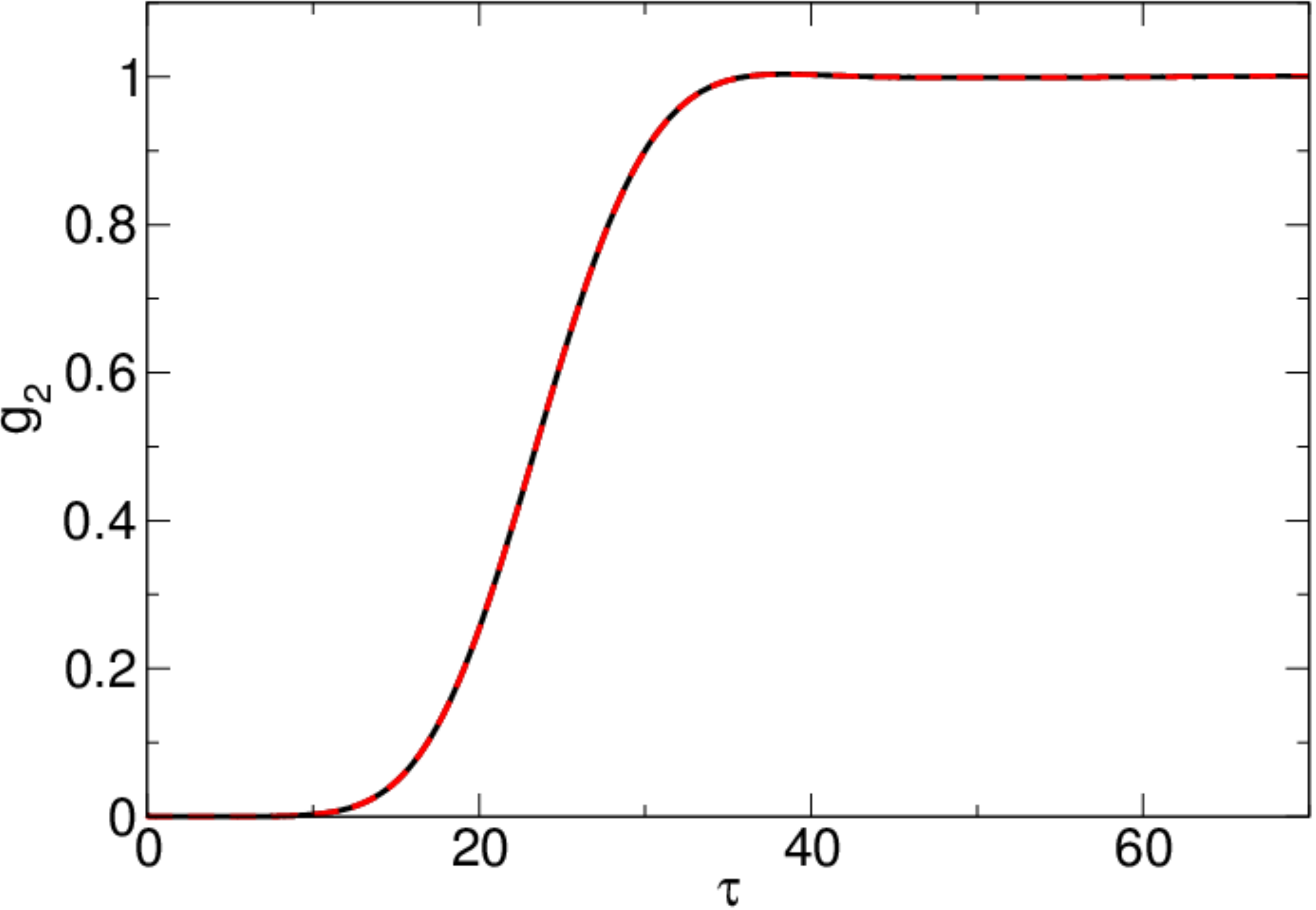}}
	\put(-0.4,14){{(a)}}
	\put(-0.4,9){{(b)}}
	\put(-0.4,4){{(c)}}
\end{picture}
\caption{(a)  single-photon (circles) and two-photon (triangles) transmission spectrum for a weak probe field, for selected values $C/2=0$ and $C/2=0.2$ of the infinite-range interaction strength. The linear transmission is independent of $C$, while the two-photon spectrum exhibits a shift in the maximum transmission by an amount $C/2$.
Other parameters: $N=200$, $\Gamma _{1D}=1, \Gamma ' =3$, $\Omega= 2$, $\delta _L= 0$,
$\mathcal{E}=10^{-6}$.
(b) Contour plot of the two-photon transmission spectrum $T_2=\langle {b}^\dagger _\text{+,out}(t){b}^\dagger _\text{+,out}(t){b}  _\text{+,out}(t){b} _\text{+,out}(t)\rangle/\mathcal{E} ^4$, as functions of interaction strength $C/2$ and two photon detuning $\Delta =\delta$ ($\delta _L=0$). Cuts of the contour plot (illustrated by the dashed lines) are plotted in a).
(c)  Comparison of $g^{(2)}(\tau)$ evaluated by numerical simulations
(red dashed line) and S-matrix theory (black line). For constant infinite range interactions
$C=1$ and $\Delta=0$, the interactions induce photon antibunching.
Other parameters: $N=20$, $\Gamma _{1D}=2, \Gamma ' =2$, $\Omega= 1$, $\delta _L= 0$, $\mathcal{E}=10^{-6}$.}
\label{I_g20_mapping:fig}
\end{figure}
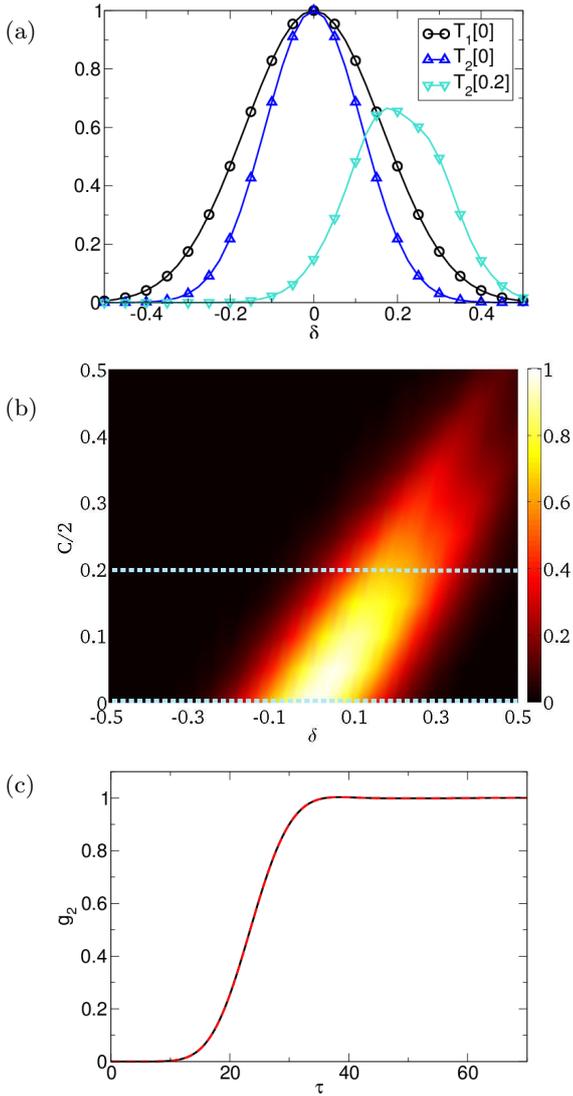
The spin model formalism can be easily extended to include arbitrary atomic interactions,
providing a powerful tool to study quantum nonlinear optical effects.
As a concrete example, we consider a system in which atoms can interact directly
over a long range, such as via Rydberg states \cite{Gorshkov_PRL11,Bienias_preprint,OTT13}
or photonic crystal bandgaps~\cite{Douglas_preprint}.
The total Hamiltonian is given by
\begin{eqnarray}
 H&=&-(\delta_L +\ii\frac{\Gamma '}{2})\sum _j a^\dagger _j a _j
-\Omega\sum _j (a^\dagger _js_j +s^\dagger _j a _j)\nonumber\\
&-& \ii\frac{\Gamma _{1D}}{2}\sum _{j,l}e^{\ii k_\text{in}|z_j-z_l|}a^\dagger _j a _l
+\frac{1}{2}\sum _{j,l}U_{ij}s^\dagger _j s^\dagger _l s _l s _j\nonumber\\
&+& H _\text{drive}
\label{ham:eq}
\end{eqnarray}
in which $U_{ij}$ represents a dispersive interaction between atoms $i$ and $j$
when they are simultaneously in state $ |s\rangle $.
As we are primarily interested
in demonstrating the use of our technique, we take here a ``toy model'' where atoms
experience a constant infinite-range interaction, $U_{ij}/2\equiv C$.
Such a case enables the numerical results to be intuitively understood,
although we note that other choices of $U_{ij}$ do not increase the numerical complexity.
In particular we are interested in studying the propagation of a constant
weak coherent input field through the atomic ensemble. The corresponding driving
then is given by
$H _\text{drive}=\mathcal{E} \sum _j(a^\dagger _j e^{\ii k_\text{in}z_j}e^{-\ii \Delta t}+a _j e^{-\ii k_\text{in}z_j}e^{\ii \Delta t})$,
where $\mathcal{E}\ll \Gamma$ is the amplitude of the constant driving field,
$\Delta=\delta-\delta _L$ the detuning from two photon resonance
condition, and the initial state is given by the global atomic ground state
$|\psi _{i}\rangle =|g \rangle^{\otimes N}$~\cite{Mollow_PRA75,Chang_NAT07}.
With infinite-range interaction, one spin flip to state $|s\rangle _{\tilde{j}}$ shifts the energies
of all other states $|s\rangle _{j}$ by an amount $C$. A second photon should then be able
to propagate with perfect transparency, provided it has a detuning compensating for
the energy shift $C$, thus ensuring the two-photon resonance condition is satisfied.
As we result, we expect to see a transparency window for two photons, whose central frequency
shifts linearly with $C$.

This predicted behavior can be confirmed by plotting the transmitted intensity
fraction,
$T_1\!\!\!\!\!\!=\!\!\!\!\! I/I_\text{in}\!\!\!\!\!=\!\!\!\!\!\langle b_\text{+,out}^\dagger(t) b_\text{+,out}(t)\rangle/\mathcal{E}^2 $,
and also the second-order correlation function
$T_2\!\!\!=\langle {b}^\dagger _\text{+,out}(t){b}^\dagger _\text{+,out}(t){b}  _\text{+,out}(t){b} _\text{+,out}(t)\rangle/\mathcal{E}^4$,
which corresponds roughly to the two-photon transmission.
Fig.~\ref{I_g20_mapping:fig} shows the single-photon transmission $T_1$
and two-photon transmission $T_2$
as a function of the interaction strength $C$ and detuning from two photon resonance $\Delta$.
As expected, $T_1$ shows a peak at $\Delta =0$ independently of the interaction intensity
$C$; instead the peak in $T_2$ shifts towards $\Delta =C/2$ with increasing $C$.
The decay of $T_2$ for increasing $C$ can be intuitively understood by noting that we have a constant coherent state input, in which photons are randomly spaced,
causing two photons to enter the medium at different times.
Thus, until the second photon enters, the first photon propagates as a single polariton
detuned by $\Delta$ from the single-photon transparency condition, getting partially
absorbed in the process.
By increasing the interaction we increase the detuning for this single polariton
and consequently its absorption, explaining the trend observed for $T_2$ in
Fig.~\ref{I_g20_mapping:fig}. A quantitative description of this phenomenon is given in Appendix D.

Field correlation functions like intensity
$I=\langle b_\text{+,out}^\dagger(t) b_\text{+,out}(t)\rangle$
or $g_{2}(\tau)=\langle {b}^\dagger _\text{+,out}(t){b}^\dagger _\text{+,out}(t+\tau){b}  _\text{+,out}(t+\tau){b} _\text{+,out}(t)\rangle/I^2$
can be computed according to the following strategy. First
we switch from Heisenberg representation to Schr\"odinger representation,
so that for the intensity we get:
$I=\langle b_\text{+,out}^\dagger(t) b_\text{+,out}(t)\rangle=\langle \psi(t)|{b}^\dagger _\text{+,out}{b} _\text{+,out}|\psi(t)\rangle$.
The time evolved wave function,
$|\psi (t)\rangle$, is determined by numerically evolving the initial spin state
$|\psi _i\rangle$ under $H$ for
a time $t$. Then, the state immediately after detection of one photon,
${b} _\text{+,out}|\psi(t)\rangle=|\phi\rangle$, is evaluated by expressing $b_\text{+,out}$
in terms of spin operators using the input-output formalism:
$b_\text{+,out}=\mathcal{E}e^{\ii k_\text{in}z_R}-\ii\Gamma _{1D}/2\sum _j\sigma _{ge}e^{\ii k_\text{in}(z_R-z_j)}$.
Finally we obtain the intensity by computing the probability
of the one-photon detected state, $I=\langle\phi|\phi\rangle$.\\

For $g_{2}(\tau)$ an extra step is needed. Its numerator describes the process of detecting two photons, the first at time $t$ and the second at
time $t+\tau$:
$f(t+\tau)=\langle {b}^\dagger _\text{+,out}(t){b}^\dagger _\text{+,out}(t+\tau){b}  _\text{+,out}(t+\tau){b} _\text{+,out}(t)\rangle$. As before we can switch to
Schr\"odinger picture,
$f(t+\tau)=\langle \psi(t)|{b}^\dagger _\text{+,out}e^{\ii H\tau}{b}^\dagger _\text{+,out}{b}  _\text{+,out}e^{-\ii H\tau}{b} _\text{+,out}|\psi(t)\rangle$, and evaluate the
state after dectection of the first photon, ${b} _\text{+,out}|\psi(t)\rangle=|\phi\rangle$.
Then detection of a second photon after a time $\tau$ entails performing
an extra evolution under $H$
and annihilating a photon, that is:
${b} _\text{+,out}e^{-\ii H\tau}|\phi\rangle={b} _\text{+,out}|\phi (\tau)\rangle$.
Finally, we evaluate the quantity
$f(t+\tau)=\langle \phi(\tau)|{b}^\dagger _\text{+,out}{b}  _\text{+,out}|\phi (\tau)\rangle$,
by again expressing $b_\text{+,out}$ in terms of spin operators.
In Fig.~\ref{I_g20_mapping:fig}c, we plot the numerically obtained result for
$g^{(2)}(\tau)$, for the case where infinite-range interactions are turned on ($C=1$) and for a weak coherent input state with detuning $\Delta=0$. In such a situation, one expects for the single-photon component of the coherent state to transmit perfectly, while the two-photon component is detuned from its transparency window and becomes absorbed. This nonlinear absorption intuitively yields the strong anti-bunching dip $g^{(2)}(\tau=0)<1$. We also evaluate this second-order correlation function using the analytical
result for the two-photon S-matrix in Eq.~(\ref{eq:S2}), which shows perfect agreement as expected.

\section{Conclusion}
We have shown that the dynamics of a few photons, propagating under strong interactions mediated by quantum emitters, is fully and efficiently characterized by the dynamics of an open quantum spin model. As the spin model is solvable by standard quantum optical techniques for open systems, our approach provides an easily implementable recipe for the exact numerical study of a large class of quantum nonlinear optical systems. As an example, it provides an attractive alternative to numerical simulations of propagation through cold Rydberg gases, where typically the continuous field is finely discretized and its degrees of freedom are explicitly kept track of~\cite{Gorshkov_PRL11}. Our technique can also be used to treat a number of strongly nonlinear systems based upon atomic saturation~\cite{CRY08}, where only approximate effective theories for field propagation were previously available. We anticipate that the availability of exact results will greatly aid in the development of effective theories for the generally challenging problem of quantum many-body states of light, and that our model will shed new insight on conceptual links between such systems and quantum spin systems.

\acknowledgments
We thank H.J. Kimble, A.V. Gorshkov, and H. de Riedmatten for stimulating discussions.
MTM acknowledges support from a La Caixa - Severo Ochoa PhD Fellowship. DEC acknowledges support from Fundacio Cellex Privada Barcelona, the Ramon y Cajal program, and the Marie Curie CIG project ATOMNANO.

TC, MTM and TS contributed equally to the work.

\appendix

\section{Scattering theory}

\subsection{Connecting S-matrix to atomic correlation functions}

We present here the full derivation of the decomposition of the S-matrix elements in terms of time-ordered atomic correlation functions. The starting point is the definition of the S-matrix in Eq.~\eqref{eq:s_1}. Since the output operators commute between themselves because of the indistinguishability of photons, they can be freely ordered by decreasing times. Introducing the time ordering operator $T$ and also using Eq.~\eqref{eq:bound_1}, the operators in Eq.~\eqref{eq:s_1} can be written as
\begin{multline}
T\big[\big(b_\text{in}(t_1)+\sqrt{\gamma}a(t_1)\big)..\big(b_\text{in}(t_n)+\sqrt{\gamma}a(t_n)\big)\big]\,\times \\
\times b^\dagger_\text{in}(t_1')..b^\dagger_\text{in}(t_n').
\end{multline}
It is natural to label the terms above by the number $m$ of system operators $a$ present in each term. Thanks to the fact that $[a(t), b_\text{in}(t')] = 0$ for $t' > t$, all the input operators can be moved to the right of the spin operators.
Thus, the term of order $m$ will be of the form
\begin{equation}
\label{eq:s_3}
\bra{0}T\big[a(t_1)..a(t_m)\big]\,b_\text{in}(t_{m+1})..b_\text{in}(t_n)b^\dagger_\text{in}(t_1')..b^\dagger_\text{in}(t_n')\ket{0},
\end{equation}
and can be simplified using the commutation relations between input operators $[b_\text{in}(t), b^\dagger_\text{in}(t')] = \delta(t-t')$. This manipulation results in a sum of $(n!)^2/(m!)^2(n-m)!$ terms for each original term of order $m$. Each term of the sum consists of $n-m$ delta functions multiplied by a correlation function of the form
\begin{equation}
\bra{0}T\big[a(t_1)..a(t_m)\big]b^\dagger_\text{in}(t_1')..b^\dagger_\text{in}(t_m')\ket{0}.
\end{equation}
Since the $a$ operators commute with the $b^\dagger_\text{in}$ operators at later times, the time ordering operator can be extended to all the operators in the correlation function.
Using again Eq.~\eqref{eq:bound_1} to express the input operators one gets
\begin{multline}
\bra{0}T\big[a(t_1)..a(t_m)\,\big(b^\dagger_\text{out}(t_1')-\sqrt{\gamma}a^\dagger(t_1')\big)..\times \\
\times \big(b^\dagger_\text{out}(t_m')-\sqrt{\gamma}a^\dagger(t_m')\big)\big]\ket{0}.
\end{multline}
However, since the operators $b^\dagger_\text{out}$ commute with all the operators $a$ on the left, only
\begin{equation}
\bra{0}T\big[a(t_1)..a(t_m)a^\dagger(t_1')..a^\dagger(t_m')\big]\ket{0}
\end{equation}
remains, which reproduces Eq.~\eqref{eq:s_7}.

\subsection{Evolution under the effective Hamiltonian}

In Eq.~\eqref{eq:s_7} the vacuum state $\ket{0}$ stands for $\left\vert 0\right\rangle _{b}\left\vert g\right\rangle ^{\otimes N}$, i.e., the vacuum state of field modes and the ground state of all the atoms. The atomic operators are in the Heisenberg picture, $a(t)=e^{iHt}ae^{-iHt}$, and $H$ is the Hamiltonian of the whole system without the driving field.
By taking the term with $t_{1}>t_{2}...>t_{m}>t_{1}^{\prime }>t_{2}^{\prime
}...>t_{m}^{\prime }$ as an example (our argument holds for any time ordering), we show now that Eq.~\eqref{eq:s_7} can be evaluated by effectively evolving system operators as $a(t)=e^{\ii H_\mathrm{eff}t}a e^{-\ii H_\mathrm{eff}t}$. The quantum regression theorem is applied here to
eliminate the bath or photonic degree of freedom, and results in%
\begin{eqnarray}
&&\left\langle 0\right\vert _{b}\left\langle 0\right\vert
a(t_{1})...a(t_{m})a^{\dagger }(t_{1}^{\prime })...a^{\dagger
}(t_{m}^{\prime })\left\vert 0\right\rangle _{b}\left\vert 0\right\rangle
\notag \\
&=&Tr[ae^{\mathcal{L}(t_{1}-t_{2})}a...a^{\dagger }e^{\mathcal{L}%
(t_{m-1}-t_{m})}a^{\dagger }\rho (0)],  \label{O}
\end{eqnarray}%
where $\rho (0)=\left\vert g\right\rangle \left\langle g\right\vert \otimes
\left\vert 0\right\rangle _{b}\left\langle 0\right\vert $ and $\mathcal{L}$
is the Lindblad super-operator of the system, defined in the main text. $%
\mathcal{L}$ contains a deterministic part, which generates an evolution driven by $H_\mathrm{eff}$ and which conserves the number of excitations, and a jump part, which reduces the number of atomic excitations. Because of the form of the correlators, which contain an equal number of atomic creation and annihilation operators, the jump part of the evolution of the operators gives a vanishing contribution to the correlation function, proving what was stated above.

\section{Examples of S-matrix elements}

In this section of the Appendix we provide explicit examples of how to use the results derived above to calculate specific matrix elements. In particular we present the case of the scattering of 1) $n$ photons on a two-level atom coupled to a one-directional waveguide and 2) one and two photons scattering on a Rydberg-EIT system.

\subsection{$n$ photons scattering on a single two-level atom}

The S-matrix element $S^{(n)}$ for the scattering of $n$ photons can be in general decomposed in a part that is a series of products of lower-order elements and a part that cannot be expressed in such a way. The latter is called fully connected part of the matrix element and physically corresponds to a $n$-body interaction, and is denoted by $\ii \mathcal{T}^{(n)}$. Thus, knowing how to decompose the S-matrix, one has to calculate $\ii \mathcal{T}^{(j)}$ with $1 \leq j \leq n$ to construct $S^{(n)}$. Furthermore, it can shown that the fully connected part in an element of order $n$ can be obtained by the system correlation function of order $n$.
We show here how to calculate such elements using the results presented above, for the case of a single two-level atom coupled to a one-directional waveguide. It will be possible to appreciate the simplicity of our formalism compared the more cumbersome method used in Ref. \cite{PLE12} to obtain the same result.

We start from the relation between the connected part of the matrix element and the correlation function of atomic operators
\begin{multline}
\label{B1}
\ii\mathcal{T}^{(n)}_ {[\mathbf{p};\mathbf{k}]} = \frac{(-\Gamma)^n}{(2\pi)^n}\int \prod_{i=1}^n\,dt_i\,dt_i'\,e^{\ii({p_i}t_i-{k_i}t'_i)} \times \\
\times\,\langle T \tilde{\sigma}_-(t_n)...\tilde{\sigma}_-(t_1)\tilde{\sigma}_+(t'_n)...\tilde{\sigma}_+(t'_1)\rangle,
\end{multline}
where $\tilde{\sigma}_-(t) = e^{\ii H_\mathrm{eff}t}\sigma_-e^{-\ii H_\mathrm{eff}t}$, with $H_\mathrm{eff} = (\omega_{eg} - \ii\Gamma/2)\sigma_{ee}$.
The time ordering in the correlator gives $2n!$ possible orderings, but it is easy to arrive to the conclusion that only orderings which start on the left with a $\sigma_-$ operator and alternate $\sigma_+$ and $\sigma_-$ give a non-zero contribution. It is also immediate to see that the number of this possible orderings is $(n!)^2$. A possible ordering is for instance
\begin{equation}
\langle\tilde{\sigma}_-(t_n)\tilde{\sigma}_+(t_n')..\tilde{\sigma}_-(t_1)\tilde{\sigma}_+(t_1')\rangle,
\end{equation}
which, expressing explicitly the time dependence, is equal to
\begin{equation}
\langle\sigma_-e^{-\ii H_\mathrm{eff}(t_n-t_n')}\sigma_+..\sigma_-e^{-\ii H_\mathrm{eff}(t_1-t_1')}\sigma_+\rangle.
\end{equation}
Since $H_\mathrm{eff}$ is diagonal in the $\ket{g},\ket{e}$ basis, we can insert identity operators between the $\sigma$ operators in the form of $\ket{g}\bra{g} + \ket{e}\bra{e}$ in order to evaluate the Hamiltonians in the exponents. We immediately end up with $\prod_{i=1}^n e^{-\ii\alpha(t_i-t_i')}$ where we have defined $\alpha = \omega_{eg} - \ii\Gamma/2$. Inserting this result in Eq.~\eqref{B1} we get
\begin{multline}
\ii\mathcal{T}^{(n)}_ {[\mathbf{p};\mathbf{k}]} = \frac{(-\Gamma)^n}{(2\pi)^n} \int^{+\infty}_{-\infty}\,dt_n\,\int^{t_n}_{-\infty}dt_n'\,...\int^{t_1}_{-\infty}dt_1'\,\bigg[\\
\prod_{i=1}^n\,e^{\ii({p_i}-\alpha)t_i}\,\,e^{-\ii({k_i}-\alpha)t'_i} + \text{permutations}\bigg],
\end{multline}
where the permutations are over the two sets of incoming and outgoing photon frequencies $k_i$ and $p_i$. The integral gives
 \begin{multline}
\label{eq:res}
\ii\mathcal{T}^n_{[\mathbf{p};\mathbf{k}]} = -2\pi\ii\bigg(\frac{\Gamma}{2\pi}\bigg)^n\,\bigg\{\prod^{n-1}_{l=1}\bigg(\sum_{i=1}^l \, \Delta_i \bigg)^{-1}\,\times\\ \times \prod_{m=1}^n\bigg(k_m - \alpha + \sum_{i=1}^{m-1}\,\Delta_i \bigg)^{-1} + \text{permutations}\bigg\} \times \\
\times \delta\big(\sum_{i=1}^n \,\Delta_i\big),
\end{multline}
where $\Delta_i = k_i-p_i$, which coincides with the result of Ref. \cite{PLE12}.

\subsection{One and two-photon scattering on a Rydberg-EIT system}

The effective Hamiltonian of the Rydberg-EIT system is given by%
\begin{equation}
H_{\mathrm{eff}}=H_{0}+H_{\mathrm{HC}}+\frac{1}{2}\sum_{ij}U_{ij}s_{i}^{%
\dagger }s_{j}^{\dagger }s_{j}s_{i},
\end{equation}%
where $H_{0}$ is given by Eq. \eqref{eit_ham:eq}, and the hardcore
interaction is%
\begin{equation}
H_{\mathrm{HC}}=\frac{U_{0}}{2}\sum_{j}(a_{j}^{\dagger }a_{j}+s_{j}^{\dagger
}s_{j})(a_{j}^{\dagger }a_{j}+s_{j}^{\dagger }s_{j}-1)
\end{equation}%
with $U_{0}\rightarrow \infty $ corresponding to the three-level atom.

For the single incident right moving photon with momentum $k$, it follows
from Eqs.~\eqref{eq:s_1} and \eqref{eq:s_7} that the $S$-matrix element is%
\begin{eqnarray}
S_{p+;k+}^{(1)} &=&\delta _{pk}-\int_{-\infty }^{+\infty }\frac{dtdt^{\prime }%
}{\pi }e^{\ii(pt-kt^{\prime })}  \notag \\
&&\times \left\langle 0\right\vert \mathcal{T}\tilde{O}_{+}(t)\tilde{O}%
_{+}^{\dagger }(t^{\prime })\left\vert 0\right\rangle,
\end{eqnarray}%
with
\begin{equation}
O_{+}=\sqrt{\frac{\Gamma _{1D}}{4}}\sum_{j}a_{j}e^{-\ii k_{\mathrm{in}}z_{j}}.
\end{equation}%
This element describes the amplitude of a single transmitted photon with momentum $p$, using the time ordered correlation function of the system operator. We set for notational simplicity the speed of light $c=1$. The second term of $S_{p+;k+}^{(1)}$, i.e., the
Fourier transform of the correlator%
\begin{eqnarray}
&&\int_{-\infty }^{+\infty }\frac{dtdt^{\prime }}{\pi }e^{\ii(pt-kt^{\prime
})}\left\langle 0\right\vert \mathcal{T}\tilde{O}_{+}(t)\tilde{O}%
_{+}^{\dagger }(t^{\prime })\left\vert 0\right\rangle  \notag \\
&=&\ii\frac{\Gamma _{1D}}{4\pi }\delta _{pk}\sum_{ij}e^{-\ii(k_{\mathrm{in}%
}+k)(z_{i}-z_{j})}[G_{0}(k)]_{ij}^{aa}
\end{eqnarray}%
can be obtained from the Green's function $G_{0}(\omega )=1/(\omega -\mathcal{H}%
_{1}) $ with the single-particle Hamiltonian%
\begin{equation}
\mathcal{H}_{1}=\left(
\begin{array}{cc}
(-\delta _{L}-\ii\frac{\Gamma ^{\prime }}{2})\delta _{ij}-\ii\frac{\Gamma _{1D}}{%
2}e^{\ii k_{\mathrm{in}}\left\vert z_{i}-z_{j}\right\vert } & \Omega \delta
_{ij} \\
\Omega \delta _{ij} & 0%
\end{array}%
\right).
\end{equation}%
Here, we have expressed $\mathcal{H}_1$ in the basis $\{\left\vert a_{i}\right\rangle ,\left\vert
s_{i}\right\rangle \}$, and $[G_{0}(k)]_{ij}^{\sigma \sigma ^{\prime }}$
denotes the element $\left\langle \sigma _{i}\right\vert G_{0}(k
)\left\vert \sigma _{j}^{\prime }\right\rangle $ with $\sigma ,\sigma
^{\prime }=a,s$. The element $S_{p+;k+}^{(1)} \equiv T_{k}\delta _{pk}$ gives rise to
the transmission coefficient%
\begin{equation}
T_{k}=1-\ii\frac{\Gamma _{1D}}{2}\sum_{ij}[G_{0}(k)]_{ij}^{aa}e^{-\ii k_{\mathrm{%
in}}(z_{i}-z_{j})}.
\end{equation}

For two right-going incident photons with momenta $k_{1}$ and $k_{2}$, the two-photon $S$-matrix element $S_{p_{1}+,p_{2}+;k_{1}+,k_{2}+}^{(2)}$ describes the
amplitude to a final state with two transmitted photons of momenta $p_{1}$ and $p_{2}$.
By Eqs.~\eqref{eq:s_1} and \eqref{eq:s_7} in the main text, we find that%
\begin{eqnarray}
S_{p_{1}+,p_{2}+;k_{1}+,k_{2}+}^{(2)} &=&T_{k_{1}}T_{k_{2}}(\delta
_{p_{1}k_{1}}\delta _{p_{2}k_{2}}+\delta _{p_{1}k_{2}}\delta _{p_{2}k_{1}})
\\
&&+\frac{\Gamma _{1D}^{2}}{4(2\pi )^{2}}\sum_{i_{1}i_{2}}%
\sum_{j_{1}j_{2}}e^{-\ii k_{\mathrm{in}}(z_{i_{1}}+z_{i_{2}})}\times  \notag \\
&&G_{i_{1}i_{2};j_{1}j_{2}}^{aa;aa}(p_{1},p_{2};k_{1},k_{2})e^{\ii k_{\mathrm{in%
}}(z_{j_{1}}+z_{j_{2}})}, \notag
\label{b13}
\end{eqnarray}%
with%
\begin{eqnarray}
&&G_{i_{1}i_{2};j_{1}j_{2}}^{aa;aa}(p_{1},p_{2};k_{1},k_{2})  \notag \\
&=&\int dt_{1}^{\prime }dt_{2}^{\prime }dt_{1}dt_{2}e^{\ii(p_{1}t_{1}^{\prime
}+p_{2}t_{2}^{\prime }-k_{1}t_{1}-k_{2}t_{2})}  \notag \\
&&\times \left\langle \mathcal{T}\tilde{a}_{i_{1}}(t_{1}^{\prime })\tilde{a}%
_{i_{2}}(t_{2}^{\prime })\tilde{a}_{j_{1}}^{\dagger }(t_{1})\tilde{a}%
_{j_{2}}^{\dagger }(t_{2})\right\rangle _{c}.
\end{eqnarray}%
Here, $\left\langle \mathcal{T}\tilde{a}_{i_{1}}(t_{1}^{\prime })\tilde{a}%
_{i_{2}}(t_{2}^{\prime })\tilde{a}_{j_{1}}^{\dagger }(t_{1})\tilde{a}%
_{j_{2}}^{\dagger }(t_{2})\right\rangle _{c}$ denotes the connected four-point Green's function, which involves an interaction between the two excitations in the system. The S-matrix also contains terms arising from disconnected correlation functions, e.g., $\left\langle \mathcal{T}\tilde{a}%
_{i_{1}}(t_{1}^{\prime })\tilde{a}_{j_{1}}^{\dagger }(t_{1})\right\rangle
\left\langle \mathcal{T}\tilde{a}_{i_{2}}(t_{2}^{\prime })\tilde{a}%
_{j_{2}}^{\dagger }(t_{2})\right\rangle $ describing the linear propagation of each excitation separately, which yields the term in \eqref{b13} proportional to $T_{k_1}T_{k_2}$.

The interaction between two excitations under the Hamiltonian $H_{\mathrm{HC}%
}+\sum_{ij}U_{ij}s_{i}^{\dagger }s_{j}^{\dagger }s_{j}s_{i}/2$ can be represented by the ladder diagram shown in Fig. 6. This diagram is represented mathematically by the two-body T-matrix, $T(E)$, which satisfies the Lippmann-Schwinger equation,
\begin{equation}
\mathbf{T}(E)=\mathbf{U}+\mathbf{U}\Pi _{0}(E)\mathbf{T}(E).  \label{T2}
\end{equation}%
Here, $E=k_1+k_2$ is the total energy, the vacuum
bubble $\Pi _{0}(E)=(E-\mathcal{H}_{2})^{-1}$ is given in terms of $\mathcal{%
H}_{2}=\mathcal{H}_{1}\otimes I_{2N}+I_{2N}\otimes \mathcal{H}_{1}$, and the
interaction matrix $\mathbf{U}$ has the diagonal element $%
U_{ij}^{aa}=U_{ij}^{as}=U_{ij}^{sa}=U_{0}$ and $U_{ij}^{ss}=U_{ij}$ in the
basis $\{\left\vert a_{i}a_{j}\right\rangle ,\left\vert
a_{i}s_{j}\right\rangle ,\left\vert s_{i}a_{j}\right\rangle ,\left\vert
s_{i}s_{j}\right\rangle \}$. The solution of Eq. (\ref{T2}) is $\mathbf{T}%
(E)=1/(\mathbf{U}^{-1}-\Pi _{0}(E))$. The S-matrix can then be written as
\begin{eqnarray}
&&S_{p_{1}+,p_{2}+;k_{1}+,k_{2}+}^{(2)}=T_{k_{1}}T_{k_{2}}\delta
_{p_{1}k_{1}}\delta _{p_{2}k_{2}}-\ii \frac{\Gamma _{1D}^{2}}{8\pi }\delta
_{p_{1}+p_{2},k_{1}+k_{2}}\times  \notag \\
&&\sum_{\substack{ iji^{\prime }j^{\prime }  \\ \sigma _{1}\sigma
_{1}^{\prime };\sigma _{2}\sigma _{2}^{\prime }}}[w^{\ast
}(p_{1},p_{2})]_{ij}^{\sigma _{1}\sigma _{1}^{\prime }}[\mathbf{T}%
(E)]_{ij;i^{\prime }j^{\prime }}^{\sigma _{1}\sigma _{1}^{\prime };\sigma
_{2}\sigma _{2}^{\prime }}[w(k_{1},k_{2})]_{i^{\prime }j^{\prime }}^{\sigma
_{2}\sigma _{2}^{\prime }}  \notag \\
&&+(p_{1}\leftrightarrow p_{2}),
\end{eqnarray}%
where%
\begin{equation}
\lbrack w(k_{1},k_{2})]_{i^{\prime }j^{\prime }}^{\sigma \sigma ^{\prime
}}=\sum_{j_{1}j_{2}}e^{\ii k_{\mathrm{in}%
}(z_{j_{1}}+z_{j_{2}})}[G_{0}(k_{1})]_{i^{\prime }j_{1}}^{\sigma
a}[G_{0}(k_{2})]_{j^{\prime }j_{2}}^{\sigma ^{\prime }a}.
\end{equation}%

\begin{figure}
\includegraphics[width=70mm]{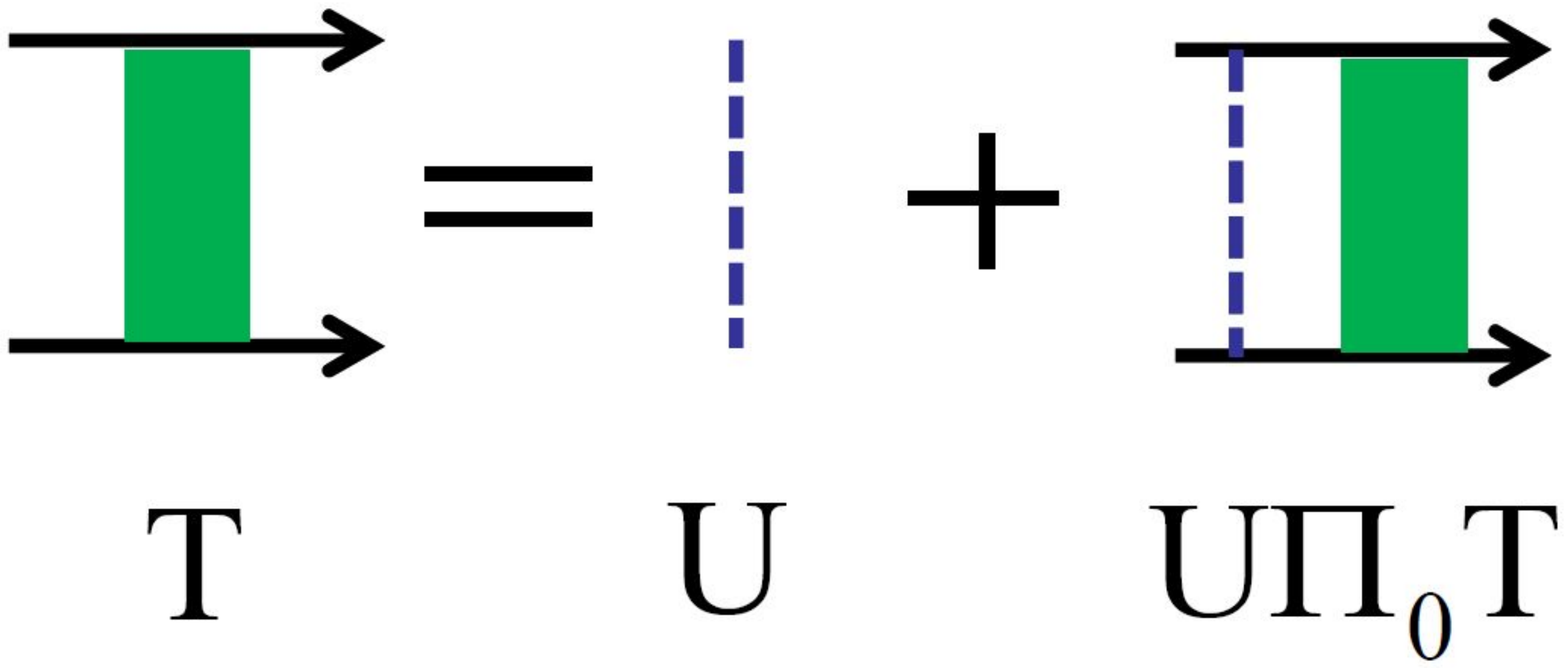}
\caption{The Feynman diagram for the $T $-matrix.}
\label{g2tr}
\end{figure}

The Fourier transform of $S_{p_{1}+,p_{2}+;k_{1}+,k_{2}+}^{(2)}$ results in the
wavefunction%
\begin{eqnarray}
&&\psi (x_{c},x)=\int \frac{dp_{1}dp_{2}}{2\pi }%
S_{p_{1}p_{2};k_{1}k_{2}}^{(2)}e^{\ii(p_{1}x_{1}+p_{2}x_{2})}  \notag \\
&=&\frac{e^{\ii Ex_{c}}}{2\pi }\{2T_{\frac{E}{2}+k}T_{\frac{E}{2}-k}\cos (kx)-\ii%
\frac{\Gamma _{1D}^{2}}{4}\sum_{\substack{ iji^{\prime }j^{\prime }  \\ %
\sigma _{1}\sigma _{1}^{\prime };\sigma _{2}\sigma _{2}^{\prime }}}%
[F(x)]_{ij}^{\sigma _{1}\sigma _{1}^{\prime }}  \notag \\
&&\times \lbrack \mathbf{T}(E)]_{ij;i^{\prime }j^{\prime }}^{\sigma
_{1}\sigma _{1}^{\prime };\sigma _{2}\sigma _{2}^{\prime }}[w(\frac{E}{2}+k,%
\frac{E}{2}-k)]_{i^{\prime }j^{\prime }}^{\sigma _{2}\sigma _{2}^{\prime
}}\},  \label{wf2}
\end{eqnarray}%
of two transmitted photons, where the relative momentum $k=(k_{1}-k_{2})/2$,
the center of mass coordinate $x_{c}=(x_{1}+x_{2})/2$, and the relative
coordinate $x=x_{1}-x_{2}$. The symmetric function%
\begin{eqnarray}
&&[F(x)]_{ij}^{\sigma \sigma ^{\prime }}=-\ii\sum_{i_{1}i_{2}}\frac{e^{-\ii k_{%
\mathrm{in}}(z_{i_{1}}+z_{i_{2}})}}{E-\varepsilon _{l}-\varepsilon
_{l^{\prime }}}\sum_{ll^{\prime }}\chi _{l}(a_{i_{1}})\chi _{l^{\prime
}}(a_{i_{2}})  \notag \\
&&\tilde{\chi}_{l}^{\ast }(\sigma _{i})\tilde{\chi}_{l^{\prime }}^{\ast
}(\sigma _{j}^{\prime })[e^{\ii(\frac{E}{2}-\varepsilon _{l^{\prime
}})x}\theta (x)+e^{-\ii(\frac{E}{2}-\varepsilon _{l})x}\theta (-x)]  \notag \\
&&+(x\rightarrow -x)
\end{eqnarray}%
is defined by the eigenstates $\left\vert \chi _{l}\right\rangle $ and $%
\left\vert \tilde{\chi}_{l}\right\rangle $ of $\mathcal{H}_{1}$ and\ $%
\mathcal{H}_{1}^{\mathrm{\dagger }}$ with the corresponding eigenenergies $%
\varepsilon _{l}$ and $\varepsilon _{l}^{\ast }$, $\left\langle \sigma
_{i}\left\vert \chi _{l}\right\rangle \right. =\chi _{l}(\sigma _{i})$ and $%
\left\langle \sigma _{i}\left\vert \tilde{\chi}_{l}\right\rangle \right. =%
\tilde{\chi}_{l}(\sigma _{i})$. Knowledge of the wave function \eqref{wf2} in the case where the two incident photons have the same momentum, $k_1=k_2=E/2$, enables one to calculate the second-order correlation function for the outgoing field, $g^{(2)}(x)=\left\vert \pi \psi
(x_{c},x)/T_{E/2}^{2}\right\vert ^{2}$.

We compare the result of $g^{(2)}(x)$ from the scattering theory and that
from numerically solving the effective spin model \eqref{ham:eq} with the weak driving
field in Fig.~\ref{I_g20_mapping:fig}c in the main text, which shows that they agree with each %
other perfectly.

\section{Generalized master equation}

In this section, we extend our derivation of the atomic master equation, in order to calculate the response of the system to an arbitrary few-photon input state (as opposed to a classical or coherent state). The
initial state is generally written as $\left\vert \psi _{\mathrm{in}%
}\right\rangle =\left\vert \varphi _{\mathrm{in}}\right\rangle \otimes
\left\vert \chi _{\mathrm{in}}\right\rangle $, where $\left\vert \chi _{%
\mathrm{in}}\right\rangle $ is the initial state of the system, and%
\begin{equation}
\left\vert \varphi _{\mathrm{in}}\right\rangle =\sum_{\{n_{v,k}\}}\varphi _{%
\mathrm{in}}(\{n_{v,k}\})\prod_{k}\left\vert \{n_{v,k}\}\right\rangle
\end{equation}%
describes an arbitrary state of incident photons in the Fock state basis by the wavefunction $\varphi _{%
\mathrm{in}}(\{n_{k}\})$. By the relation%
\begin{equation}
\left\vert n\right\rangle =\lim_{J\rightarrow 0}\frac{1}{\sqrt{n!}}\frac{%
\partial ^{n}}{\partial J^{n}}\left\vert J\right\rangle
\end{equation}%
between the Fock state $\left\vert n\right\rangle $ and the coherent state $\left\vert J\right\rangle =\sum_{n}J^{n}\left\vert
n\right\rangle /\sqrt{n!}$, we rewrite $\left\vert \varphi _{\mathrm{in}%
}\right\rangle =\mathcal{F}\left\vert \{J_{v,k}\}\right\rangle $ by the
operator%
\begin{equation}
\mathcal{F}=\lim_{\{J_{v,k}\}\rightarrow 0}\sum_{\{n_{k}\}}\varphi _{\mathrm{%
in}}(\{n_{v,k}\})\prod_{k}\frac{1}{\sqrt{n_{v,k}!}}\frac{\partial ^{n_{v,k}}%
}{\partial J_{v,k}^{n_{v,k}}}
\end{equation}%
acting on the coherent state $\left\vert \{J_{k}\}\right\rangle $.

The evolution of the reduced density matrix $\rho _{s}(t)=Tr_{b}[\mathcal{U}%
(t)\rho (0)\mathcal{U}^{\dagger }(t)]$ is determined by%
\begin{equation*}
\mathcal{U}(t)=\mathcal{T}\exp [-\ii\int_{0}^{t}dsH(s)],
\end{equation*}%
where $\rho (0)=\left\vert \psi _{\mathrm{in}}\right\rangle \left\langle
\psi _{\mathrm{in}}\right\vert $ is the density matrix of the initial state,
and $H(t)$ is the total Hamiltonian in the rotating frame with $%
b_{v,k}\rightarrow b_{v,k}e^{-\ii vkt}$. By the relation $\left\vert \varphi _{%
\mathrm{in}}\right\rangle =\mathcal{F}\left\vert \{J_{v,k}\}\right\rangle $,
the reduced density matrix reads%
\begin{equation}
\rho _{s}(t)=\mathcal{F}^{\ast }\mathcal{F}Tr_{b}[\mathcal{U}(t)\left\vert
\{J_{v,k}\}\right\rangle \left\langle \{J_{v,k}\}\right\vert \rho _{s}(0)%
\mathcal{U}^{\dagger }(t)],
\end{equation}%
where $\rho _{s}(0)=\left\vert \chi _{\mathrm{in}}\right\rangle \left\langle
\chi _{\mathrm{in}}\right\vert $ is the initial density matrix of the
system. The displacement transformation $V_{\mathrm{d}}\left\vert
\{J_{k}\}\right\rangle =e^{\frac{1}{2}\sum_{v,k}\left\vert
J_{v,k}\right\vert ^{2}}\left\vert 0\right\rangle $ leads to%
\begin{equation}
\rho _{s}(t)=\mathcal{F}^{\ast }\mathcal{F}e^{\sum_{v,k}\left\vert
J_{v,k}\right\vert ^{2}}\rho _{J}(t),  \label{re}
\end{equation}%
and the generating density matrix%
\begin{equation}
\rho _{J}(t)=Tr_{b}[\mathcal{U}_{\mathrm{d}}(t)\left\vert 0\right\rangle
\left\langle 0\right\vert \rho _{s}(0)\mathcal{U}_{\mathrm{d}}^{\dagger
}(t)],
\end{equation}%
where the unitary transformation%
\begin{equation}
\mathcal{U}_{\mathrm{d}}(t)=\mathcal{T}\exp [-\ii\int_{0}^{t}dsH_{\mathrm{d}%
}(s)]
\end{equation}%
is given by $H_{\mathrm{d}}(s)=V_{\mathrm{d}}H(s)V_{\mathrm{d}}^{\dagger }$.
The Hamitonian $H_{\mathrm{d}}(s)=H(s)+H_{J}(s)$ is obtained by replacing
the operator $b_{v,k}$ by $b_{v,k}+J_{v,k}$ in the Hamiltonian $H(s)$, where
$H_{J}(s)$ describes the system under the driving fields $J_{v,k}$. In the
density matrix $\rho _{J}(t)$, the initial state becomes the vacuum state due to
the displacement transformation, thus the evolution of $\rho _{J}(t)$
satisfies the master equation%
\begin{equation}
\partial _{t}\rho _{J}(t)=\mathcal{L}\rho _{J}(t)-\ii[H_{J}(t),\rho _{J}(t)],
\end{equation}%
where $\mathcal{L}$ is the Lindblad super-operator of the system without $%
H_{J}$.

In conclusion, Eq. (\ref{re}) establishes the relation bewteen the evolution
of the reduced density matrix $\rho _{s}(t)$ for a few incident photons (which could be a non-classical state) and
the evolution of the reduced density matrix $\rho _{J}(t)$ for the system
with classical (coherent state) driving fields $J_{v,k}$. As a result, the
few photon scattering problem can be understood as the perturbation
expansion of the driving strength $J_{v,k}$.

\section{Linear approximation for two photon transmission $T_2$}
\begin{figure}
\includegraphics[width=90mm,angle=0,clip]{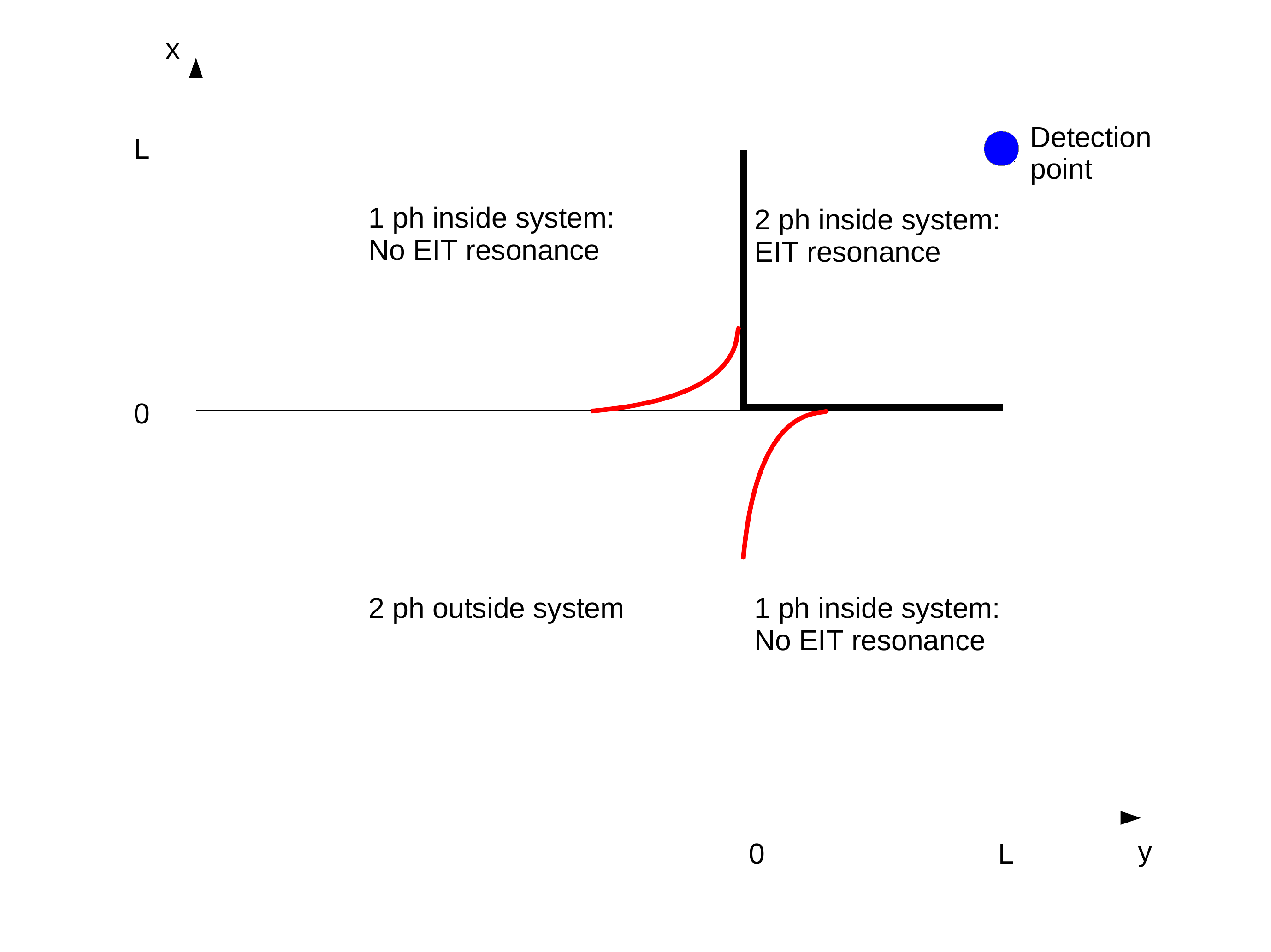}
\caption{An intuitive explanation of how the two-photon wave function evolves within the atomic medium
(located in the region $0<x,y<L$) can be found by considering a larger space, where first the two photons
are both outside the medium ($x,y<0$) and where one photon enters the medium first.
We specifically consider the case where infinite-range interactions cause the two photons to align with the
EIT transparency condition when they are both inside the medium, as discussed further in the main text.
The two-photon transmission is proportional to the value of the two-photon wave function at $x,y=L$.}
\label{toy_model:fig}
\end{figure}

In this section of the Appendix we provide an intuitive explanation based on linear optics
to the behavior of the two-photon transmission $T_2$
depicted in Fig.~\ref{I_g20_mapping:fig}a,b.
By indicating with $x$ and $y$ the coordinates of the first and second photons, we
can divide the space into four regions according to their positions: both photons outside
the atomic medium, one photon inside and one outside, and both photons inside the medium, see Fig.~\ref{toy_model:fig}.
As we are studying an infinite-range interaction, within each region the evolution is effectively linear. The different dispersion
relations in each region, however, leads to non-trivial boundary conditions at their edges. Relevant to our discussion is the
value of the two-photon wave function at the boundaries $0<x<L$ and $0<y<L$. In particular, evolution in the region
$0<x, y<L$ with these boundary conditions dictates the two-photon transmission, which is proportional to the value of
the two-photon wave function at $(x,y)=(L,L)$.\\
We study the case where the two photons have detunings $\Delta=C/2$. In this case, infinite-range
interactions cause these two photons to satisfy the EIT transparency condition when both photons are
 inside the medium. Then, there are two qualitatively different regimes for the boundary values of the
 two-photon wave function, depending on the detuning from two-photon
resonance $\delta$ (for simplicity we assume $\delta _L=0$, so that $\Delta=\delta-\delta _L=\delta$):
the small detuning regime, $\delta < \Delta _{EIT}$, in which a single photon can travel with high transmission
through the medium
and reach the detection point; and the large detuning regime, $\delta > \Delta _{EIT}$,
in which a single photon is absorbed and cannot reach the detection point.

\begin{figure}[top]
\setlength{\unitlength}{1cm}
\begin{picture}(8,10)
	\put(0.55,5) {\includegraphics[width=67mm,angle=0,clip]{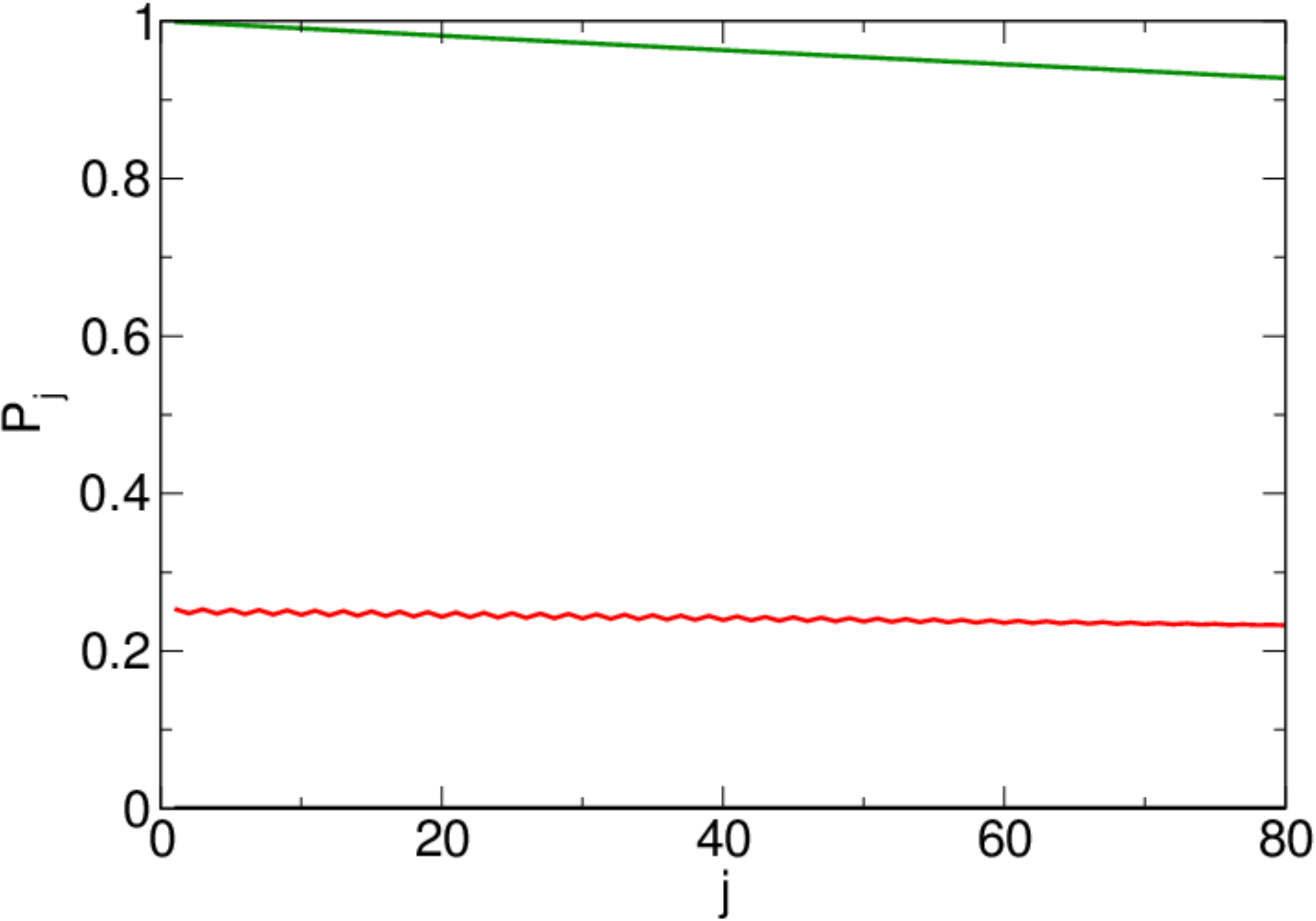}}	
	\put(0.55,0) {\includegraphics[width=67mm,angle=0,clip]{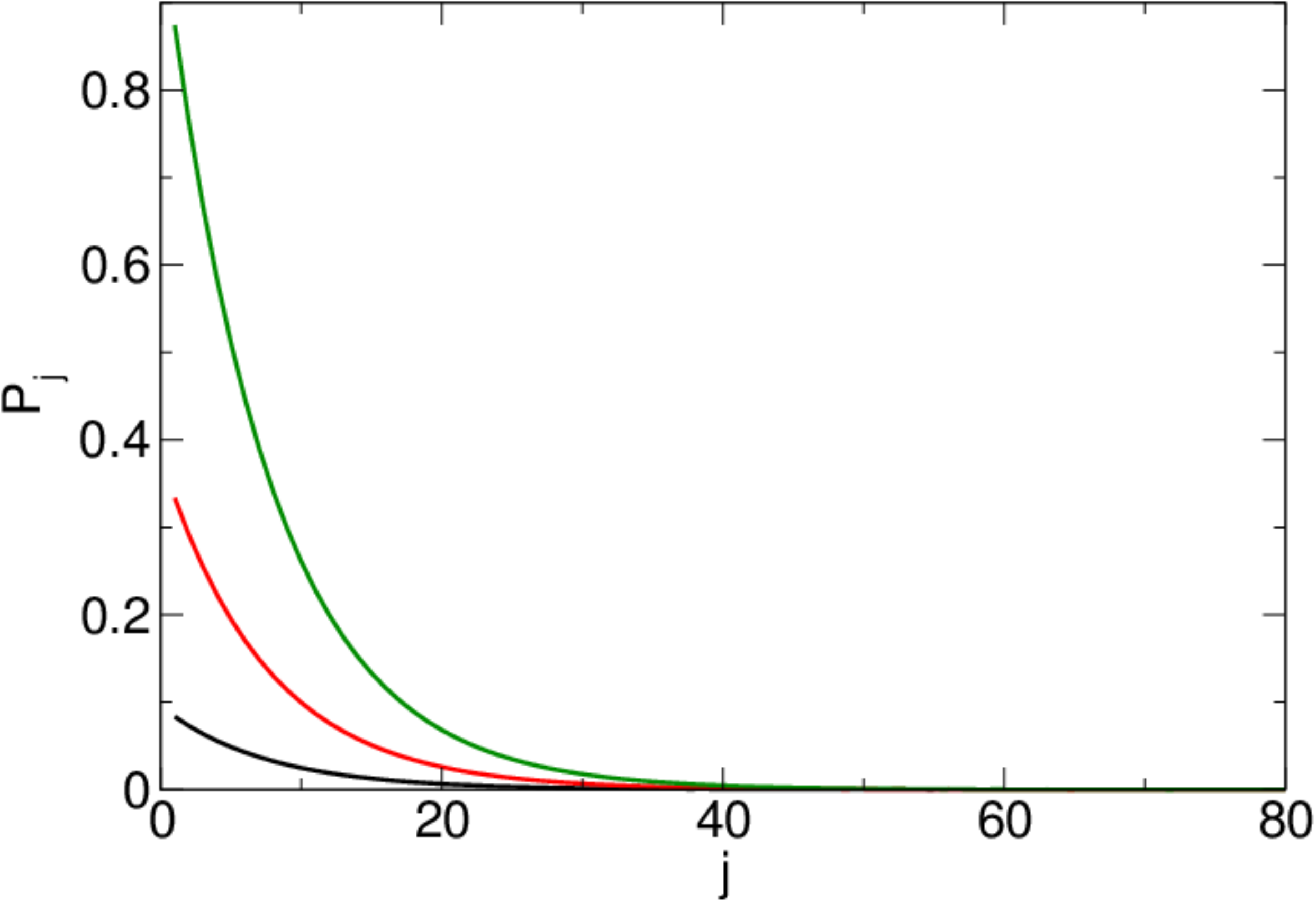}}
	\put(0,9){{(a)}}
	\put(0,4){{(b)}}
\end{picture}
\caption{Single photon probability $\langle b^\dagger _{\rm out}(jd)b_{\rm out}(jd)\rangle$ (green)
and single polariton probabilities $\langle\sigma _{ee} ^j\rangle$ (black) and
$\langle\sigma _{ss}^j\rangle$ (red) as function of atom position $j$, for asymptotically large
time at $\Delta=\delta=0.1$ (a) and at $\Delta=\delta=1$ (b). Probabilities are in unit of
$\alpha ^2$. Other parameters: $N=80, \Omega =2, \Gamma=3,\Gamma _{1D}=1, \alpha=10^{-6}$.}
\label{lin_intensity:fig}
\end{figure}

\begin{figure}
\includegraphics[width=65mm,angle=0,clip]{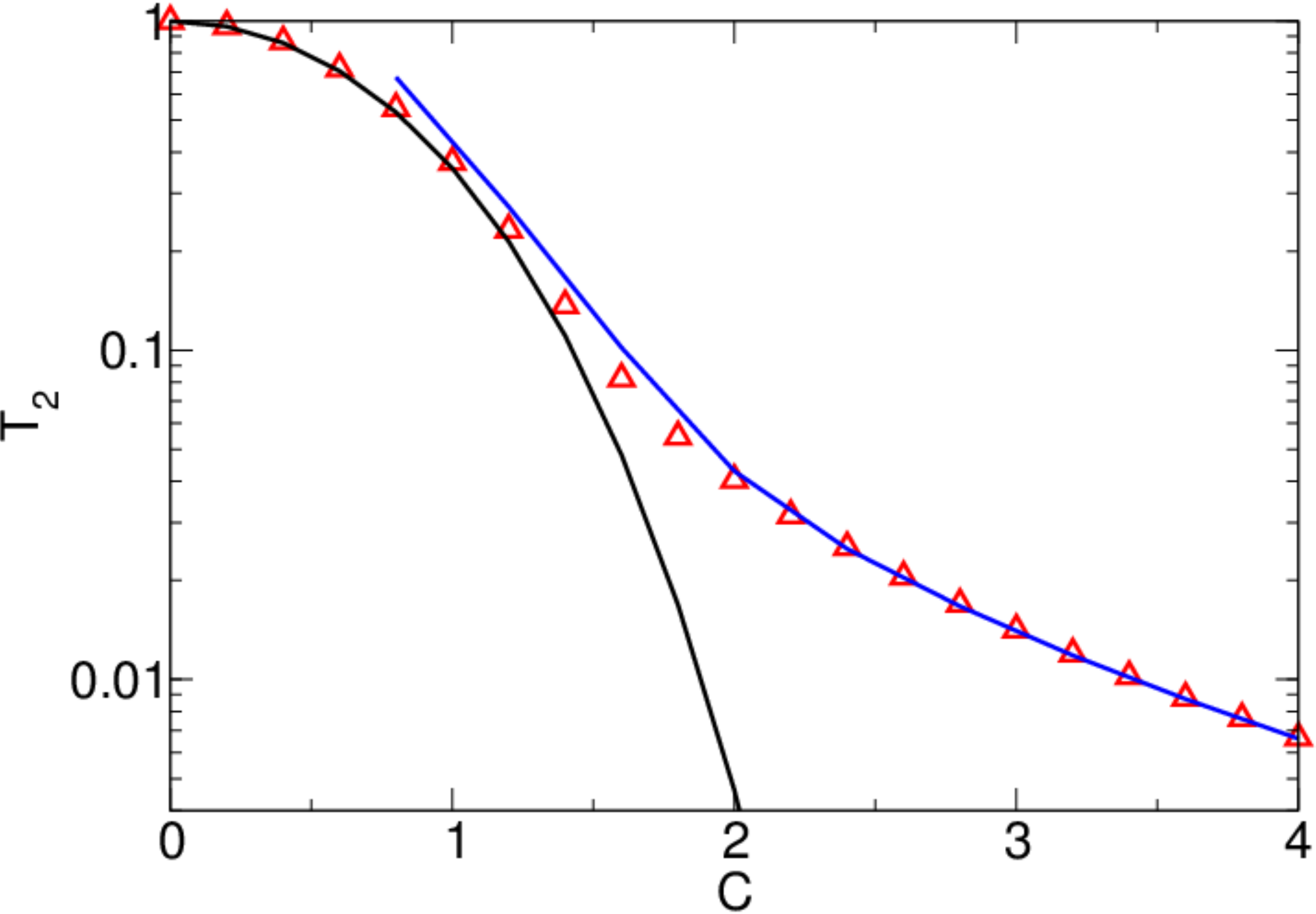}
\caption{$T_2$ versus $C=2\delta$: comparison of the full simulation $T_2$ (red triangles) with the
simple models for the small detuning regime (where one expects $T_2=\sqrt{T_1}$, black line) and the large
detuning regime (blue curve), where the two-photon transmission is obtained from the propagation
of a single photon localized at the front of the medium (near $z=0$).}
\label{comparison:fig}
\end{figure}

\emph{Small detuning regime.---} Within the first regime $\delta=C/2 < \Delta _{EIT}$,
a single polariton still fits within the EIT transparency window and exhibits high transmission
through the medium. A representative probability distribution for a single excitation in this regime is
illustrated in Fig.~\ref{lin_intensity:fig}a. Now, we consider the two-excitation manifold.
After detection of the first outgoing photon,
the second photon can be at any position inside the ensemble with nearly uniform probability, and
on average has to travel over a distance $Nd/2$ before reaching the detection point.
Therefore we expect $T_2=|\exp (-\ii k(\delta)Nd/2 )|^2=\sqrt{T_1}$. The
predicted behaviour is in good agreement with full simulations results
as shown in Fig.~\ref{comparison:fig} for small detuning.

\emph{Large detuning regime.---} In this regime a single photon is strongly absorbed.
This is illustrated in Fig.~\ref{lin_intensity:fig}b, where one observes a strong decay of the
single-polariton probability and field intensity inside the medium.
Within the context of Fig.~\ref{toy_model:fig}, this localized single excitation serves as the boundary
condition on the segments $0<x<L$ and $0<y<L$. Furthermore, the evolution in the region $0<x,y<L$ is effectively linear.
Thus, the effect of this boundary condition at the detection point (a two-particle problem) can be mapped onto
a simpler problem, wherein one studies how a single excitation, initialized in the shape of the localized photon,
transmits through the medium.
In this regime, we therefore can estimate that $T_2$ is twice the value of the maximum transmission
associated with this single localized photon, whose evolution we calculate numerically.

The good agreement of our simplified estimates with full simulations is demonstrated
in Fig.~\ref{comparison:fig}, in which $T_2$ as a function of $C=2\delta$ is compared for
the different cases.
%
%

\bibliographystyle{apsrev}

\begin{thebibliography}{99}

\bibitem{NLO14}
D.~E. Chang, V. Vuletic and M.~D. Lukin, Nat. Photonics \textbf{8},
  685 (2014).

\bibitem{K08}
H.~J. Kimble, Nature \textbf{453}, 1023 (2008).

\bibitem{HAR06}
M.~J. Hartmann, F.~G. S.~L. Brandao and M.~B. Plenio, Nature Phys.
  \textbf{2}, 849 (2006).

\bibitem{GRE06}
A.~D. Greentre, C. Tahan, J.~H. Cole and L.~C.~L. Hollenberg,
  Nature Phys. \textbf{2}, 856 (2006).

\bibitem{CRY08}
D.~E. Chang, \emph{et al.},
  Nature Phys. \textbf{4}, 884 (2008).

\bibitem{OTT13}
J. Otterbach, M. Moos, D. Muth, and M. Fleischhauer, Phys. Rev. Lett. \textbf{111},
  113001 (2013).

\bibitem{TUR95}
Q.~A. Turchette, C.~J. Hood, W. Lange, H. Mabuchi, and H.~J. Kimble,
Phys. Rev. Lett. \textbf{75}, 4710 (1995).

\bibitem{BIR05}
K.~M. Birnbaum, \emph{et al.}, Nature \textbf{436}, 87 (2005).

\bibitem{VET10}
E. Vetsch, \emph{et al.},
  Phys. Rev. Lett. \textbf{104}, 203603 (2010).

\bibitem{GOB12}
A. Goban, \emph{et al.}, Phys. Rev. Lett. \textbf{109}, 033603 (2012).

\bibitem{BAJ09}
M. Bajcsy, \emph{et al.}, Phys. Rev. Lett. \textbf{102}, 203902 (2009).

\bibitem{GOB14}
A. Goban, \emph{et al.}, Nature Commun. \textbf{5}, 3808 (2014).

\bibitem{rydberg_experiments}
 J.~D. Pritchard, \emph{et al.}, Phys. Rev. Lett. \textbf{105}, 193603 (2010);
 J. Pritchard, K. Weatherill, and C. Adams, Nonlinear optics using cold Rydberg atoms in
 \emph{Annual Review of Cold Atoms and Molecules} (eds K.~W.~Madison, Y. Wang, A.~M. Rey, K. Bongs) {\bf 1}, 301-350 (World Scientific, 2013);
 T. Peyronel, \emph{et al.}, Nature {\bf 488}, 57 (2012);
 O. Firstenberg, \emph{et al.}, Nature {\bf 502}, 71 (2013);
 S. Baur, D. Tiarks, G. Rempe and S. Durr, Phys. Rev. Lett. \textbf{112}, 073901 (2014)

\bibitem{LOO13}
A.~F. van Loo, \emph{et al.}, Science \textbf{342}, 1494 (2013).

\bibitem{HOI13}
I.-C. Hoi, \emph{et al.},
  Phys. Rev. Lett. \textbf{111}, 053601 (2013).

\bibitem{GAD95}
C.~W. Gardiner and M.~J. Collett, Phys. Rev. A \textbf{31}, 3761 (1985).

\bibitem{DEU95}
I.~H. Deutsch, R.~J.~C. Spreeuw, S.~L. Rolston and W.~D. Phillips,
Phys. Rev. A \textbf{52}, 1394 (1995).

\bibitem{DEC12}
D.~E. Chang, L. Jiang, A.~V. Gorshkov and H.~J. Kimble, New\ J.\ Phys. \textbf{14}, 063003
  (2012).

\bibitem{SHI09}
T. Shi and C.~P. Sun, Phys. Rev. B  \textbf{79}, 205111 (2009).

\bibitem{FAN10}
S. Fan, S.~E. Kocabas and J.-T. Shen, Phys. Rev. A \textbf{82},
063821 (2010).

\bibitem{SHI11}
T. Shi, S. Fan, C.~P. Sun, Phys. Rev. A \textbf{84}, 063803 (2011).

\bibitem{PLE12}
M. Pletyukhov and V. Gritsev, New\ J.\ Phys. \textbf{14}, 095028 (2012).

\bibitem{LUM13}
K. Lalumiere \emph{et al.}, Phys. Rev. A \textbf{88}, 043806 (2013).

\bibitem{meystre}
Meystre P., Sargent III M.
\emph{Element of Quantum Optics}, (Springer, New York, 1999).

\bibitem{eit_papers} M. Fleischhauer and M. D. Lukin, Phys. Rev. A {\bf 65}, 022314 (2002);
	M. Fleischhauer, A. Imamoglu, and J. P. Marangos, Rev. Mod. Phys. {\bf 77}, 633 (2005).

\bibitem{opticalDepth} Actually the formula $D=N(2\Gamma _{1D}/\Gamma)$ is derived
	in the limit $\Gamma'\gg\Gamma _{1D}$, where for a two-level atom the transmission on resonance is given by
	$t=1-(\Gamma _{1D}/\Gamma)\simeq \exp(-\Gamma _{1D}/\Gamma)$.
	Such a regime of high transmission per single atom is relevant for the majority of
	EIT experiments and indeed, as shown in the main text, known results of EIT
	are reproduced correctly.
	However in the opposite limit $\Gamma '\ll\Gamma _{1D}$ a single atom behaves
	like a perfect mirror~\cite{Chang_NAT07} ($t\simeq 0$) and the correct theoretical
	description requires a different expansion for the
	transmission coefficient, $t=1-(\Gamma _{1D}/\Gamma)\simeq \Gamma '/\Gamma _{1D}$.

\bibitem{Mollow_PRA75} B. R. Mollow, Phys. Rev. A {\bf 12}, 1919 (1975).

\bibitem{Chang_NAT07} D. E. Chang, A. S. S\o rensen, E. A. Demler, and M. D. Lukin, Nature {\bf 3}, 807 (2007).

\bibitem{Chang_NJP11} D. E. Chang, A. H. Safavi-Naeini, M. Hafezi, and O. Painter, New J. Phys. {\bf 13}, 023003 (2011).

\bibitem{perfectMirror} In the complementary case where $k_{\rm in}d=m\pi$, reflections from atoms constructively interfere and atomic mirrors can be realized as shown in Ref.~\cite{DEC12}.

\bibitem{Gorshkov_PRL11} A.~V. Gorshkov, J. Otterbach, M. Fleischhauer, T. Pohl and M.~D. Lukin,
Phys. Rev. Lett. {\bf 107}, 133602 (2011).

\bibitem{Bienias_preprint} P. Bienias, \emph{et al.}, arXiv:1402.7333.

\bibitem{Douglas_preprint} J. Douglas, \emph{et al.}, arXiv:1312.2435.


\end{thebibliography}

\end{document}